\documentclass[useAMS,usenatbib]{mn2e}
\usepackage{graphicx}

\newcommand{\be}{\begin{equation}}
\newcommand{\ee}{\end{equation}}
\newcommand{\bea}{\begin{eqnarray}}
\newcommand{\eea}{\end{eqnarray}}

\title[NoSOCS in SDSS. IV]{NoSOCS in SDSS. IV. The Role of Environment Beyond the 
Extent of Galaxy Clusters}
\author[Lopes et al.]{P. A. A. Lopes$^{1}$\thanks{E-mail: 
paal05@gmail.com}, A. L. B. Ribeiro$^{2}$, S. B. Rembold$^{2,3}$\\
$^{1}$Observat\'orio do Valongo, Universidade Federal do Rio de Janeiro, 
Ladeira do Pedro Antônio 43,\\ 
Rio de Janeiro, RJ, 20080-090, Brazil\\
$^{2}$Laborat\'orio de Astrof\'isica Te\'orica e Observacional -- Departamento 
de Ci\^encias Exatas e Tecnol\'ogicas\\
Universidade Estadual de Santa Cruz, 45650-000, Ilh\'eus, BA, Brazil\\
$^{3}$Universidade Federal de Santa Maria -- 97105-900, Santa Maria-RS, Brazil}
\begin{document}

\date{Accepted  Received ; in original form }

\pagerange{\pageref{firstpage}--\pageref{lastpage}} \pubyear{2008}

\maketitle

\label{firstpage}

\begin{abstract}
We are able to extend the investigation of the color-morphology-density-radius 
relations, for bright and faint galaxies, to $R \ga 3 \times R_{200}$ and to very 
low density regions, probing the transition region between cluster and field 
galaxies, and finding a smooth variation between these two populations.
We investigate the environmental variation of galaxy properties 
(and their relations), such as color, spectral type and concentration. 
Our sample comprises 6,415 galaxies that were previously selected as 
cluster members from 152 systems with $z \le 0.100$. This sample is
further divided in complete subsamples of 5,106 galaxies with 
$M_r \le M^*+1$ (from clusters at $z \le 0.100$) and 1,309 galaxies with 
$M^*+1 < M_r \le M^*+3$ (from objects at $z \le 0.045$). We characterize the 
environment as a function of the local galaxy density and global cluster 
related parameters, such as radial distance, substructure, X-ray luminosity 
and velocity dispersion. For a sample of field galaxies we also trace their 
environmental dependence using a local galaxy density estimate. Our main findings 
are: (i) The fraction of discs is generally higher than the ones for blue and 
star-forming galaxies, indicating a faster transformation of color and 
star-formation compared to structural parameters. (ii) Regarding the distance to 
the cluster center we find a small variation in the galaxy populations outside 
the virial radius. Once within that radius the fractions of each population change 
fast, decreasing even faster within $R \sim 0.3 \times R_{200}$. (iii) We also find 
a small increase in the fraction of blue faint galaxies within 
$R \sim 0.4 \times R_{200}$, before decreasing again to the most central bin. 
(iv) Our results do not indicate a significant dependence on cluster mass, 
except for the disc fraction in the core of clusters. (v) The relations 
between galaxy properties also point to no dependence on cluster mass, except for 
the scatter of the color stellar mass relation. Our results corroborate a scenario 
on which pre-processing in groups leads to a strong evolution 
in galaxy properties, before they are accreted by large clusters.
\end{abstract}


\section{Introduction}

Galaxies in the local Universe can be classified in two broad types, according to their
structure or properties related to their star formation history. Elliptical and S0 
galaxies tend to be redder, to be bulge-dominated and show little star formation, being 
called early-type galaxies. Spirals are bluer, show high star formation rate (SFR)
and are termed late-type. The presence of these two main types have a strong correlation
with environment, with the former being predominant in the most dense regions of the 
Universe (central parts of groups and clusters of galaxies), while the latter are more 
common in low-density environments, called field \citep{oem74, dre80}.
Those results can arise from differences in the intrinsic galaxy formation process 
(galaxy properties also show strong dependence on stellar mass). Another natural 
interpretation of this morphology-density relation is that galaxies are transformed 
as they move from the field to the central parts of clusters.
Several different processes acting within groups and clusters could be important 
for the transformation of late-type into early-type galaxies as they move from low 
to high density regions. These include ram-pressure stripping, strangulation, mergers 
and tidal stripping from the cluster 
potential \citep{too72, cal93, gun72, aba99, bos06}. The open 
question regarding all these mechanisms is to 
know where they become relevant, as we consider infalling galaxies into groups or 
clusters. As these systems' boundaries are hard to place it is nearly impossible to 
mark the transition between the field and groups/clusters.

Recent work has shown that the variation of galaxy types extend to very low density 
regions where the cluster influence should me minimal \citep{mat04, bal04, bah13}. 
These results became 
possible after the very large projects such as the 2dF Galaxy Redshift Survey (2dFGRS) 
and Sloan Digital Sky Survey (SDSS) became available. According to these works
there is a critical density for which the connection between SFR and environment first 
takes place. As this critical density is compatible with regions outside the cluster 
virial radius, it is suggested that galaxies are first processed in groups 
\citep{bal04}. Besides this pre-processing \citet{bah13} also 
suggests that {\em overshooting} (\citealt{lud09}; another indirect mechanism) 
is responsible for transforming galaxies, as well as ram pressure due to low density 
gas in the outskirts of clusters.

The morphology-density (and color-density) relation is known to have a strong dependence 
on mass, with red early-type objects being the most massive. That agrees with the 
scenario of hierarchical growth of structure, 
from which we expect the most massive halos to be 
in the most dense parts of the Universe. As a consequence we expect 
the morphology-density relation to be a morphology-halo mass relation. However, results
in the literature are contradictory. For instance, \citet{del12} and \citet{val11} 
find no dependence on the fraction of early-type
galaxies (bulge-dominated and red galaxies), and properties of the cluster 
red-sequence (RS), with parent halo mass. On the other hand, \citet{cal12} 
find a small dependence between morphology and halo mass. 
These differences may be associated to different halo 
mass ranges. The work of \citet{del12} is based on a narrow cluster mass 
range ($14 \le log M_H/h^{-1} M_{\odot} \le 14.8$), using simulated data, while the 
results of \citet{val11} are restricted to clusters with 
$500 < \sigma_P < 1100$ km s$^{-1}$. \citet{cal12} consider systems 
spanning a much wider mass range ($12 \le log M_H/h^{-1} M_{\odot} \le 15$). Another
recent work \citep{hoy12} based on even more massive systems than
\citet{cal12} ($13 \le log M_H/h^{-1} M_{\odot} \le 15.8$; but a 
different lower mass cut) find no significant variation of galaxy type fraction with
halo mass.

This work is the fourth of a series aiming to investigate cluster and
galaxies' properties at low redshifts ($z \le 0.1$). Our main goal is 
to study galaxy properties in a variety of environments,
from the central parts of groups/clusters to the field. 
For that we obtain the variation
of the fractions of galaxy populations (traced by different properties) with 
local and global environment, characterized by the local galaxy density, 
distance to the center of the parent cluster, 
cluster mass (traced by the velocity dispersion and X-ray luminosity) 
and substructure. Besides investigating the environmental variation of the galaxy
populations we check its dependence on galaxy cluster mass. We also investigate the 
differences in the field and group/cluster galaxies and the transition between
such populations. All the analysis is performed on two different luminosity ranges 
($M_r \le M^*+1$ and $M^*+1 < M_r \le M^*+3$), so that we can assess the impact of 
environment for giant and dwarf galaxies separately. The environmental influence 
is also investigated in different ranges of galaxy stellar mass.

This paper is organized as follows: $\S$2 has the data description, where we 
also discuss the local galaxy density estimates, the field sample and the 
separation of the galaxy populations. In $\S$3 we present the environmental 
variation of different galaxy populations according to different environment 
tracers. In $\S$4 we show how the relations between different galaxy properties 
depend on environment. In $\S$5 we summarize and discuss our main results. 
The cosmology assumed in this work considers $\Omega_{\rm m}=$0.3, 
$\Omega_{\lambda}=$0.7, and H$_0 = 100$ $\rm h$ $\rm km$ $s^{-1}$ Mpc$^{-1}$, 
with $\rm h$ set to 0.7. For simplicity, in the following we are going to use 
the term ``cluster'' to refer loosely to groups and clusters of galaxies, except 
where explicitly mentioned.

\section{Data}

In the first paper of this series (hereafter paper I, \citealt{lop09a}) 
we defined a cluster sample from the supplemental version of the Northern Sky 
Optical Cluster Survey (NoSOCS, \citealt{lop03, lop04}). For that we used data
from the 5th Sloan Digital Sky Survey (SDSS) release. This sample comprises 
7,414 systems well sampled in SDSS DR5 (details in paper I). NoSOCS has its origin 
on the digitized version of the Second Palomar Observatory Sky Survey (POSS-II; 
DPOSS, \citealt{djo03}). In \citet{gal04} and \citet{ode04} the photometric calibration 
and object classification for DPOSS, respectively, are described. The 
supplemental version of NoSOCS \citep{lop04} goes deeper ($z \sim 0.5$), but 
covers a smaller region than the main NoSOCS catalog \citep{gal03, gal09}.

For a subset of the 7,414 NoSOCS systems with SDSS data we extracted a 
subsample of  low redshift galaxy clusters ($z \le 0.100$). This subsample
comprises 127 clusters, for which we had enough spectra in SDSS for 
spectroscopic redshift determination, as well as to select cluster members and 
perform a virial analysis, obtaining estimates of velocity dispersion, 
physical radius and mass ($\sigma_P$, $R_{500}$, $R_{200}$, $M_{500}$ and 
$M_{200}$; details in paper I). This low-redshift sample was 
complemented with more massive systems
from the Cluster Infall Regions in SDSS (CIRS) sample (\citealt{rin06},
hereafter RD06). CIRS is a collection of $z \le 0.100$ X-ray selected clusters 
overlapping the SDSS DR4 footprint. The same cluster parameters listed above 
were determined for these 56 CIRS clusters. 

In the second paper of this 
series (hereafter paper II, \citealt{lop09b}) we investigated the scaling 
relations of clusters using this combined sample of 183 clusters at 
$z \le 0.100$, except for three systems that are not used for having  
biased values of $\sigma_P$ and mass due to projection effects. These 3 
clusters are Abell 1035B, Abell 1291A, and Abell 1291B. We plan to investigate 
the properties of these interacting clusters in detail in a future work. 
Details about this
low-redshift sample and the estimates obtained for the clusters can be
found in papers I and II. For the clusters with at least five galaxy members within 
R$_{200}$ we also have a substructure estimate, based on the DS, or $\Delta$ 
test \citep{dre88}. Details about this test can be found in paper I.

The redshift limit of the sample ($z = 0.100$) is due to incompleteness in
the SDSS spectroscopic survey for higher redshifts, where galaxies fainter 
than $M^*+1$ are missed, biasing the dynamical analysis (see discussion in section 
4.3 of \citealt{lop09a}). We eliminated interlopers using the ``shifting gapper'' 
technique \citep{fad96}, applied to all galaxies with spectra available within a 
maximum aperture of 2.50 h$^{-1}$ Mpc. We also estimated X-ray luminosity ($L_X$, 
using ROSAT All Sky Survey data), optical luminosity ($L_{opt}$) and richness 
(N$_{gals}$, \citealt{lop09a, lop09b}). The centroid of each NoSOCS cluster is a 
luminosity weighted estimate, which correlates well with the X-ray peak 
(see \citealt{lop06}).

In \citet{rib13} (hereafter paper III) we investigated the connection between
galaxy evolution and the dynamical state of galaxy clusters, indicated by their
velocity distributions. Here we focus on the investigation 
of the variation of galaxy properties (and their relations),
such as color, spectral class, stellar mass, 
and concentration, with the environment. Note
that we only use galaxies that are selected as cluster members by the
``shifting gapper'' technique. 

For the current work, we have also implemented 
one modification to the ``shifting gapper''
technique, regarding the radial offset between two consecutive galaxies. In previous 
papers of this series, if
this quantity was greater than 1.0 Mpc the ``shifting gapper'' procedure was
stopped. Then, all galaxies with a radial distance from the cluster centre 
greater or equal to this galaxy's radial distance were eliminated as 
interlopers (see \citealt{lop09a}). This maximum offset between consecutive galaxies 
is now set to be 0.75 $\times sep_{biw}$, where $sep_{biw}$ is the biweight estimate of 
the typical separation of the inner galaxies (within 0.50 h$^{-1}$ Mpc). This 
modification was motivated by the fact that for a few systems the total
number of members (N$_{tot}$) was much larger than the number of members within 
R$_{200}$ (N$_{200m}$). These were small mass groups that have
a large cluster nearby (after $\sim$ 2.0 Mpc). With the new version of this code
some lower mass systems end up not having a minimum of ten group members selected
to run the virial analysis. Due to that our sample is reduced from 180 groups and
clusters to 152, for which we have 6,415 galaxies, being 5,106 with $M_r \le M^*+1$ 
(from clusters at $z \le 0.100$) and 1,309 galaxies with 
$M^*+1 < M_r \le M^*+3$ (from objects at $z \le 0.045$). Our clusters span 
the range $150 \la \sigma_P \la 950$ km s$^{-1}$, or the equivalently in terms 
of mass, $10^{13} \la M_{200} \la 10^{15} M_{\odot}$.

\subsection {Local Galaxy Density Estimates}

When comparing literature results regarding the environmental dependence of galaxy
properties one central issue is the definition of environment. That can be local 
(associated to the galaxy neighborhood) or global (related to the large scale 
structure). The galaxy environment measure is further complicated by different
selection criteria, issues related to the survey geometry, as well as luminosity
limits. \citet{bla07} investigate the impact of group environment 
on different scales, concluding that not only the surrounding density matters.
\citet{wil10} present a multiscale model-independent 
approach to measure the galaxy density. Their method is used to investigate the
variation of the galaxy ($u-r$) color distribution (with -21.5 $\le M_r \le$ -20) on
multiscale density. \citet{mul12} compare the results from twenty 
different published environment definitions. They classify the methods to measure 
environment in two groups, those based on nearest neighbours and those using 
fixed apertures. They conclude that the local environment is best probed by 
nearest neighbours methods, while the large-scale environment is best measured by
apertures.

Considering the results from \citet{mul12} we decide to adopt a 
nearest neighbour method to estimate the local environment. The authors also mention
that the choice of $n$ - the rank of the density-defining neighbour - is very 
important, as the environment measure may loose power
in the case $n$ is larger than the number of galaxies residing in the halo.
Hence, we chose to work with the $\Sigma_5$ local galaxy density estimator, as
$n = 5$ is typically smaller than the number of galaxies we have per cluster and
is a common estimate in the literature. To estimate local galaxy densities we proceed 
as follows. For every galaxy member we compute its projected distance, d$_5$, to the 
5th nearest galaxy found around it, within a maximum 
velocity offset of 1000 $km$ $s^{-1}$ 
(relative to the velocity of the galaxy in question). The  local  density 
$\Sigma_5$  is simply given by 5/$\pi$d$_N^{2}$, and is measured in  units of 
galaxies/Mpc$^2$. Density estimates are also obtained relatively only to galaxies 
brighter than a fixed luminosity range, which we adopt as $M^* + 1.0$. On what regards
the global environment we consider the distance to the center of the parent cluster
and its mass (traced by the X-ray luminosity and velocity dispersion). We also 
inspect possible differences relative to the dynamical state of clusters, indicated
by the substructure estimate.

\subsubsection{Correction for the fiber collision issue}

The fiber collision issue affects the SDSS spectroscopic sample and the derived
density estimates. Due to a mechanical restriction spectroscopic fibers cannot  
be placed closer than 55 arcsecs on the sky.  An algorithm used for  target  
selection randomly  chooses  which  galaxy  gets a  fiber, in case of a conflict
\citep{str02}. This  problem is reduced by spectroscopic plate
overlaps, but   fiber  collisions   still   lead   to  a   $\sim$ 6\%
incompleteness in  the main galaxy  sample. Our  approach to  fix this problem is 
similar to  the one adopted by \citet{ber06} and 
\citet{lab10}.
For galaxies brighter than $r =$ 18 with no redshifts we  assume the redshift of the 
nearest neighbour  on the sky (generally the galaxy it collided with). This may 
result in some nearby galaxies to be placed at high redshift, artificially increasing
their  estimated  luminosities. Due to that the collided  galaxies  also
assume the magnitudes of their nearest neighbours, resulting in an unbiased 
luminosity distribution. Notice that the fraction of fixed galaxies is quite small 
(at  most  $6\%$, at highest densities); the above correction procedure has been 
shown to accurately match the multiplicity function of groups in mock 
catalogues \citep{ber06}; and the velocity distribution (relative to group 
centre) of the original and fixed samples are consistent \citep{lab10}.
Hence, the local galaxy density estimates take in account the fiber collision issue.
For every galaxy we want to estimate density, either group member or 
belonging to the field (see next section) in the original spectroscopic sample, we
use the fixed sample to find d$_5$ and estimate $\Sigma_5$.

\subsubsection{Field sample}

Besides computing local galaxy densities to group/cluster members we also do that
for a galaxy field sample. This sample is constructed as follows. From the whole DR7
data set we select galaxies that would not be associated to a group or cluster, 
considering the cluster catalog from \citet{gal09}. To be conservative, we consider a 
galaxy to belong to the \emph{field} if it is not found within 4.0 Mpc and not having a 
redshift offset smaller than 0.06 of any cluster from \citet{gal09}. We select more
than 60,000 field galaxies, but work with a smaller subset (randomly chosen) 
for which we derived local density estimates. 
In the end we use 2,986 field galaxies at $z \le$ 0.100 
with $M_r \le M^*+1$,  and 1,829 at $z \le$ 0.045 with $M^*+1 < M_r \le M^*+3$ (the 
same order of cluster galaxies). For these objects we compute density estimates 
in the same way as done for the cluster members.

\subsection{Absolute Magnitudes and Colors}

For the current work we consider the stacked properties of cluster and field 
galaxies (in this case, regarding local density only). We do that considering the 
radial offset (in units of $R_{200}$), absolute magnitudes, colors and
local densities of all member galaxies coming from the 152 clusters (or the field). 
Our sample consists of 5,106 bright member galaxies with $M_r \le M^*+1$ (at 
$z \le 0.100$), 1,309 faint members with $M^*+1 < M_r \le M^*+3$ (at $z \le 0.045$), 
2,986 bright field galaxies ($M_r \le M^*+1$, $z \le 0.100$) and 1,829 faint field
galaxies ($M^*+1 < M_r \le M^*+3$, $z \le 0.045$).

To compute the absolute magnitudes of each galaxy (in $ugri$ bands) we consider the 
following formula: $M_x = m_x - DM - kcorr - Qz$ ($x$ is one of the four SDSS bands
we considered), where DM is the distance modulus (considering the redshift of each 
galaxy), $kcorr$ is the k$-$correction and $Qz$ ($Q = -1.4$, \citealt {yee99}) 
is a mild evolutionary 
correction applied to the magnitudes (for each galaxy redshift). The k$-$corrections 
are obtained directly from the SDSS database, for every object in each band. 
Rest-frame colors are also obtained for all objects.

\subsection{Separation of Galaxy Populations}

Galaxies can generally be separated in two groups according to their structure and
star formation activity. Early-type (E and S0) galaxies are redder, have little star 
formation and high concentration, while late-type galaxies are bluer, show active 
star formation and are less concentrated. Here we use three galaxy parameters to
separate galaxies in two populations and investigate the variation of the fraction
of each population (specified by each of the three parameters) with environment. 
We use the ($u-r$) color, the concentration 
index $C$ (defined as the ratio of the radii enclosing 90 per cent 
and 50 per cent of the galaxy light in the r-band, $R_{90}/R_{50}$) and 
a spectral classification called $e_{class}$ (which is based on a PCA analysis of 
the SDSS spectrum database). We call blue/red
galaxies those with ($u-r$) values smaller/greater 
than 2.2 \citep{str01}. Galaxies
are considered to have low-concentration (and are called 
disc-dominated objects) if they have $C < 2.6$ (bulge-dominated galaxies 
have values greater than 2.6). The parameter $e_{class}$ ranges from about 
$-$0.35 to 0.5 for early- to late-type galaxies (or passive to star-forming), 
with the former having values typically below $-$0.05 (spectra with 
lower-temperature absorption features). 
Late-type objects have spectra with higher temperature features or emission lines.
Note that the separation value for $C$ is in agreement to what was adopted by 
\citet{str01} and \citet{kau04}.

\begin{figure}
\begin{center}
\leavevmode
\includegraphics[width=3.5in]{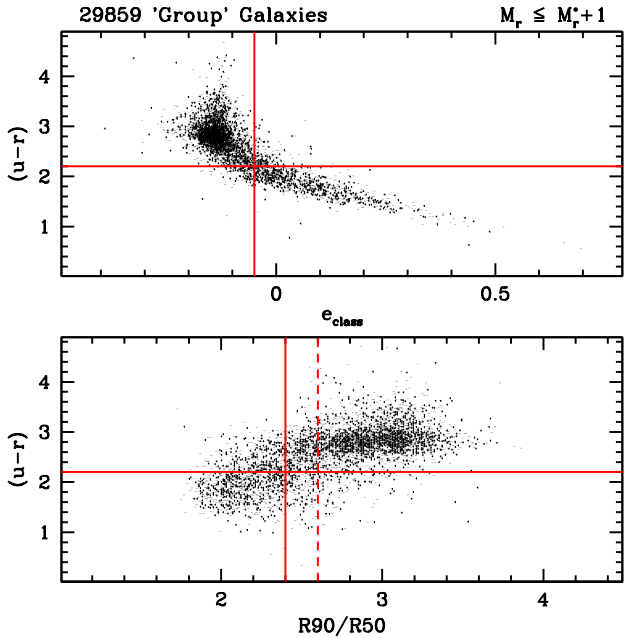}
\end{center}
\caption{The correlation between ($u-r$) color with the parameters $e_{class}$ and 
$C = R_{90}/R_{50}$ for bright galaxies ($M_r \le M^*+1$) in group/clusters regions. The
horizontal lines indicate the color separation, while the solid vertical lines 
indicate the median value of galaxies with ($u-r$) $\sim 2.2$. In the bottom panel 
the dashed vertical line shows the actual value adopted to split galaxies according
to concentration ($C = 2.6$).}
\label{fig:gals_sep1}
\end{figure}

\begin{figure}
\begin{center}
\leavevmode
\includegraphics[width=3.5in]{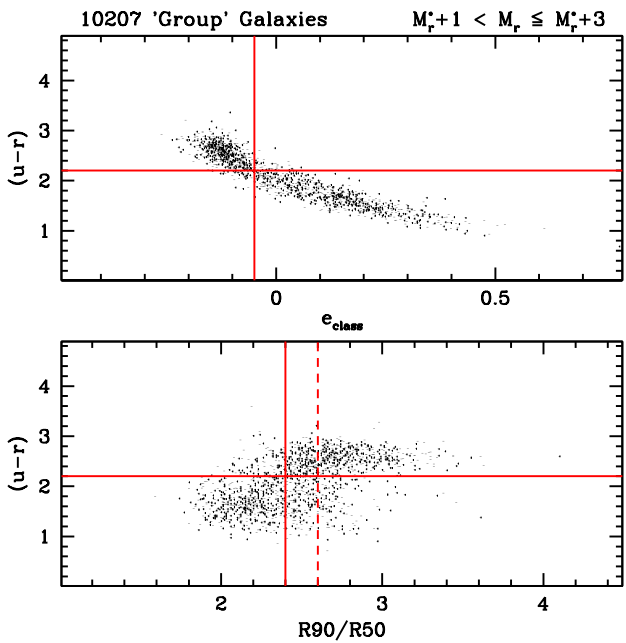}
\end{center}
\caption{The same as in the previous figure, but for faint galaxies 
($M^*+1 < M_r \le M^*+3$). The horizontal and vertical lines are the same.}
\label{fig:gals_sep2}
\end{figure}

In Fig.~\ref{fig:gals_sep1} we show the correlation between ($u-r$) color with 
$e_{class}$ and $C$ for bright galaxies ($M_r \le M^*+1$) in the groups/clusters 
regions (\emph{i.e.}, all galaxies, members and interlopers). The
horizontal lines indicate the color separation, while the solid vertical 
lines indicate the median values (for $e_{class}$ and $C$) of galaxies 
with ($u-r$) $\sim 2.2$. In the bottom panel the dashed vertical line shows 
the actual value adopted to split galaxies regarding concentration ($C = 2.6$).
Fig.~\ref{fig:gals_sep2} is analogous, but for faint galaxies 
($M^*+1 < M_r \le M^*+3$). As we can see from these two figures, for the data used 
here, $C = 2.4$ gives the median value of galaxies with ($u-r$) $\sim 2.2$. We 
consider $C = 2.6$ for concordance with the literature (actually some authors 
even consider higher values of $C$). Note also that we show only galaxies in
cluster regions and not from the whole sky; and concentration does not provide
a sharp cut as color or $e_{class}$. Our choice $C = 2.6$ provides a good compromise
between completeness and concentration of the two broad populations (early and late
type galaxies). A smaller value of $C$ would increase the completeness of early-type
objects at the cost of a larger contamination by late-types and vice-versa (see 
discussion in \citealt{str01}). 

Note that this choice of $C = 2.6$ is good enough to split galaxies in two broad 
populations (early and late type galaxies), as is the goal of next section ($\S$3).
In $\S$4 we investigate the environmental dependence of the relations between 
different galaxy properties. In that case, we analyze the relations for different
subsamples according to concentration. There we decided to consider three ranges
of concentration. We call low concentration galaxies or {\it classical discs} objects 
with $C \le 2.4$, while we consider {\it classical bulges} galaxies with 
$C > 2.8$. Objects with $2.4 < C \le 2.8$ are called intermediate types. In that
case we have purer samples of bulges and discs (avoiding contamination from the other 
type) and consider an intermediate population that can give clues to galaxy 
transformation regarding environment and stellar mass.

In the next section we investigate the environmental variation of 
the fractions of blue, high $e_{class}$ (we call those star-forming) and 
low concentration galaxies (we call those disc-dominated), which we refer as 
$F_B, F_L$ and $F_D$, respectively. Instead of separating galaxy populations using
only one parameter ({\it {e.g.}} color) we work with these three parameters 
as ``red'' does not necessarily imply ``bulge'' (or high-concentration), 
due to differences in the relations between color and morphology with density.

\section{The Environmental Variation of Blue, High $e_{class}$ and Low 
Concentration Galaxies}

In this section we investigate the variation of galaxy properties (given by fractions
of populations) with the environment, characterized by the local galaxy density, 
normalized radial distance to the parent cluster, presence of substructure, 
as well as its X-ray luminosity and velocity dispersion. In $\S$3.5 the 
environmental influence is also investigated in different ranges of galaxy stellar 
mass. When considering the local galaxy density we use cluster members and 
field galaxies. For the remaining tracers of environment we use 
only cluster members. 

\subsection{Dependence on Local Galaxy Density}

\begin{figure}
\begin{center}
\leavevmode
\includegraphics[width=3.5in]{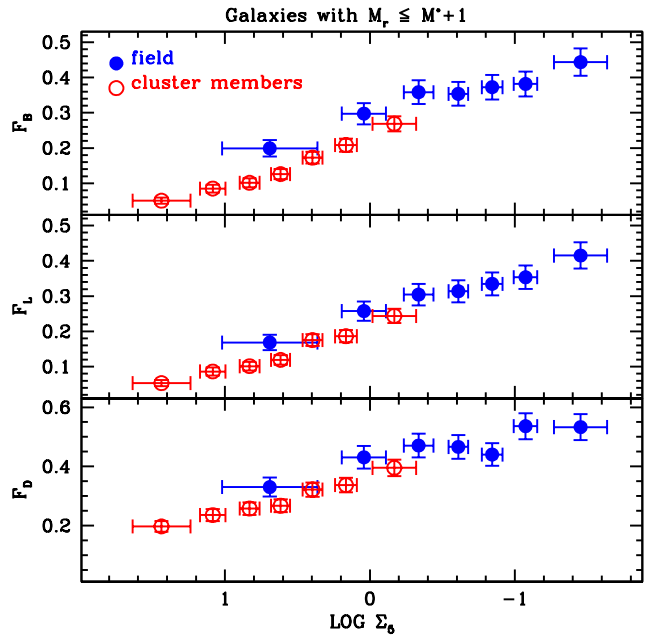}
\end{center}
\caption{From top to bottom we show the variation of $F_B, F_L$ and $F_D$ with
local galaxy density ($\Sigma_5$). Red open circles are for cluster members 
and blue filled circles for field galaxies. These results consider bright 
galaxies ($M_r \le M^*+1$).}
\label{fig:denbins}
\end{figure}

\begin{figure}
\begin{center}
\leavevmode
\includegraphics[width=3.5in]{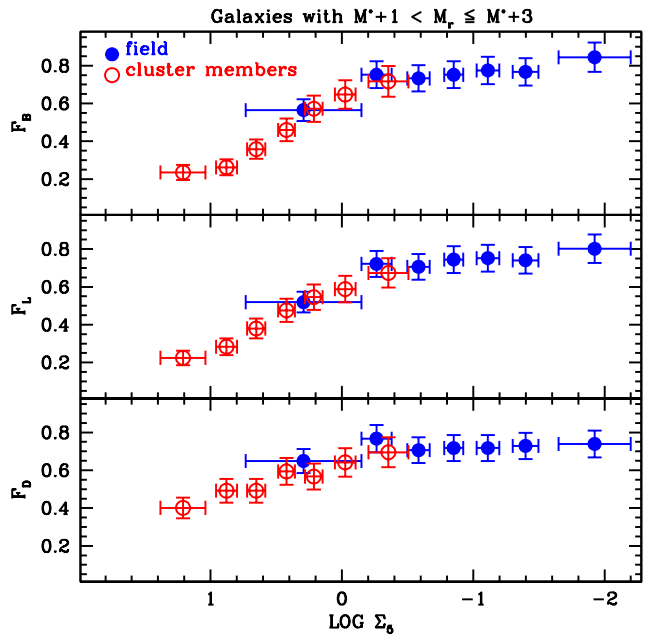}
\end{center}
\caption{Same as previous figure, but showing results for faint galaxies 
($M^*+1 < M_r \le M^*+3$).}
\label{fig:denbins2}
\end{figure}

In Fig.~\ref{fig:denbins} we show how the fractions of blue ($F_B$), 
high $e_{class}$ (or star-forming, $F_L$), and low concentration 
(or disc-dominated, $F_D$) galaxies depend on
local galaxy density ($\Sigma_5$), for the bright regime ($M_r \le M^*+1$). 
$F_B$ is shown in the top panel, while $F_L$ is in the middle panel and $F_D$ 
in the bottom. Red open circles show cluster members, while blue filled 
circles indicate field galaxies. Fig.~\ref{fig:denbins2} is analogous, 
but showing faint galaxies 
($M^*+1 < M_r \le M^*+3$).

From both figures we can see the results of the 
color- and morphology-density relations, with 
a small fraction of blue and late-type galaxies (characterized by $e_{class}$ 
and $C$) in the most dense regions of the Universe. These fractions grow 
continuously to the lower density regime. However, the pace is reduced when 
we leave clusters and reach the field for bright galaxies, while for faint 
objects the fractions actually become approximately constant. For instance, 
for bright objects we have $F_B \sim 5\%$ and $\sim 27\%$ at the highest and 
lowest density bins, for cluster members, respectively. For field galaxies we 
go from $F_B \sim 20\%$ to $\sim 44\%$ from the highest to lowest
density bins. For bright cluster members we still have $F_L \sim 5\%$ 
and $\sim 24\%$, and $F_D \sim 20\%$ and $\sim 40\%$, in the highest and 
lowest density bins, respectively. The equivalent fractions to field galaxies 
are $F_L \sim 17\%$ and $\sim 42\%$, and $F_D \sim 33\%$ and $\sim 53\%$.
The fact that the variation with density still holds to the lowest density bins 
in the field, for bright galaxies, indicates that other factors not directly 
related to clusters may act to affect these galaxies (such as
tidal effects from neighbour galaxies).

Note that in the transition between clusters and the field we might have some 
contamination on both sides; \emph{i.e.}, field galaxies that are erroneously
classified as cluster members and possibly cluster galaxies that are wrongly called 
field objects (the former would belong to systems not listed in \citealt{gal09}).
However, we do not expect these contaminations to be high as the ``shifting gapper'' 
technique \citep{fad96} was extensively tested elsewhere (with its member list
being well compared to other methods; see \citealt{lop09a}); and the cluster
catalog of \citet{gal09} is complete for rich systems in the local Universe. So,
the expected smooth transition from clusters to the field is not affected in our
results.

For bright galaxies (Fig.~\ref{fig:denbins}) we can see that the highest 
density field galaxies show larger fractions ($F_B, F_L$ and $F_D$) than cluster 
galaxies at similar density, indicating that even the cluster low density 
environment plays a role on further diminishing the fraction of blue and 
star-forming bright galaxies \citep{bah13}. For faint 
objects (Fig.~\ref{fig:denbins2}) that is not true, as at the same density 
cluster and field galaxies show the same proportion of blue and 
star-forming objects. We 
also notice that for the highest density bins the fractions ($F_B$ and $F_L$) 
do not become smaller than $\sim 20\%$ for faint galaxies, while it 
is $\sim 5\%$ for the bright regime (for $F_D$ the corresponding values are 
$\sim 40\%$ and $\sim 20\%$). That could be due to the formation of small 
(low luminosity) galaxies in the central parts of clusters, from tidal debris 
of galaxy mergers or as consequence of ram pressure stripping of larger galaxies
\citep{mas05, rib13}.

Three other interesting features stand out in Fig.~\ref{fig:denbins2}. First, there is
nearly no environmental variation for faint field objects, except for the most dense
bin. That indicates the factors acting on bright field 
galaxies are not equally effective 
for the faint objects. Second, the environmental variation is present in all bins 
for cluster members (red open circles); indicating the effectiveness of the cluster 
environment even in the outskirts. The third point is also seen in 
Fig.~\ref{fig:denbins}. $F_B$ and $F_L$ show very similar behavior to each other,
while $F_D$ is generally higher, especially for cluster objects.
That goes in line with other results in the literature suggesting that color and 
star-formation are affected much faster than structural parameters 
\citep{hom06, got03, bam09, ski09, bun10}.

\subsection{Dependence on Cluster Radius}

\begin{figure}
\begin{center}
\leavevmode
\includegraphics[width=3.5in]{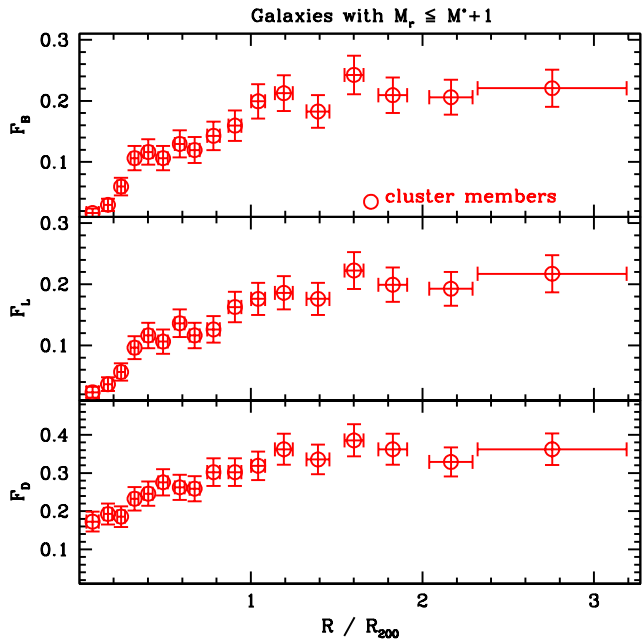}
\end{center}
\caption{From top to bottom we show the variation of $F_B, F_L$ and $F_D$ with
normalized distance to the cluster center ($R/R_{200}$). Only cluster members are 
shown. These results consider bright galaxies ($M_r \le M^*+1$).}
\label{fig:radbins}
\end{figure}

\begin{figure}
\begin{center}
\leavevmode
\includegraphics[width=3.5in]{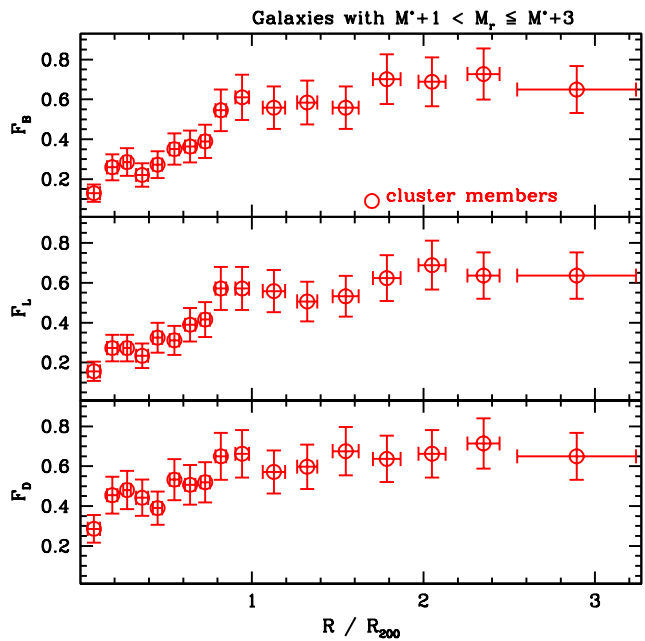}
\end{center}
\caption{Same as previous figure, but showing the variations for faint galaxies 
($M^*+1 < M_r \le M^*+3$).}
\label{fig:radbins2}
\end{figure}

Figures ~\ref{fig:radbins} and ~\ref{fig:radbins2} are analogous to the ones shown in 
the section above, but now exhibiting the variations of $F_B, F_L$ and $F_D$ with
the normalized distance to the parent cluster ($R/R_{200}$). The results
are only shown to cluster members.  Fig.~\ref{fig:radbins} is for bright and
Fig.~\ref{fig:radbins2} is for faint galaxies.

The color-morphology-radius imprint is clearly seen in both figures. Differently than 
what is seen for cluster galaxies as a function of $\Sigma_5$ (red open circles) the 
values of $F_B, F_L$ and $F_D$ stop growing (or grow much slower) after a 
given radius ($\sim R_{200}$). It is also possible to devise another abrupt change
at another radius ($\sim 0.3 \times R_{200}$). From $\sim R_{200}$ to the central part
of clusters the decline in $F_B, F_L$ and $F_D$ becomes steeper within 
$\sim 0.3 \times R_{200}$ (in agreement with \citealt{got03}). 
For bright galaxies, in the most central bin 
we have almost no blue and high $e_{class}$ ($\la 2\%$) galaxies, but the fraction 
of low $C$ objects is $\sim 17\%$. $F_B$ and $F_L$ grow to $\la 16\%$ to 
roughly the virial radius (approximated by $R_{200}$). Outside the virial radius they 
grow to $\sim 20\%$, remaining nearly constant after that. $F_D$ grows to 
$\sim 30\%$ at $\la R_{200}$, being $\la 36\%$ for larger radii. In other words, the 
population of bright galaxies do not change significantly for objects falling into 
a cluster until they are within its virial radius. Within the cluster the population
changes faster when in the core ($\sim 0.3 \times R_{200}$). That indicates the tidal 
effects (due to the cluster potential) and the interaction to the ICM (denser within 
$R_{200}$) are very effective on transforming the galaxy populations, which would
also lead to the so called anemic spirals \citep{van76}.

A similar trend is seen for faint galaxies, but different regimes may take place. 
From large to small radius it seems that $F_B, F_L$ and $F_D$ are roughly constant
until $\sim 1.8 \times R_{200}$ ($\sim 70\%$ for $F_B$), falling to a lower plateau 
between $\sim 1.8 \times R_{200}$ and $\sim R_{200}$ ($\sim 55\%$ for $F_B$) and showing
a steep decline until $\sim 0.4 \times R_{200}$ ($F_B$ decreases to $\sim 22\%$). In
the very central region (within $\sim 0.4 \times R_{200}$) there is a slight rise
in the values of $F_B, F_L$ and $F_D$ before reaching the minimum value in the most
central bin. If that rise in the central part is real it is probably associated to
the formation of small galaxies in the core of clusters, from tidal debris 
or due to ram pressure stripping of larger galaxies \citep{mas05, rib13}. 
The variation in the four regimes mentioned above
for $F_L$ and $F_D$ is similar to $F_B$, being slightly noisier for $F_D$.

Against local density $F_B, F_L$ have a minimum at $\sim 5\%$ and $\sim 20\%$ for bright
and faint objects, respectively (the corresponding values are $\sim 20\%$ and $\sim 40\%$
for $F_D$). Regarding the physical radius the minimum values of $F_B, F_L$ are 
$\sim 2\%$ and $\sim 14\%$ for bright and faint objects, respectively (for $F_D$ 
those values are $\sim 17\%$ and $\sim 29\%$). Within $R \sim R_{200}$, for faint objects,
and at all radius for bright galaxies, it can also be seen
that $F_D$ is generally higher than $F_B$ and $F_L$, indicating that
infalling galaxies have their color and spectral properties transformed quicker 
than their structure, corroborating the results from previous section.

One final result regarding radius deserves our attention. As seen in 
\citet{bah13} the variation of galaxy types reach far 
distances from the cluster centers. Here we show this variation to be real, 
for bright and faint galaxies, up to $\ga 3 \times R_{200}$, suggesting that 
pre-processing by groups, {\em overshooting} and possibly ram pressure in 
the outskirts of those systems act to transform the galaxy population.

\subsection{Dependence on Cluster Substructure}

\begin{figure}
\begin{center}
\leavevmode
\includegraphics[width=3.5in]{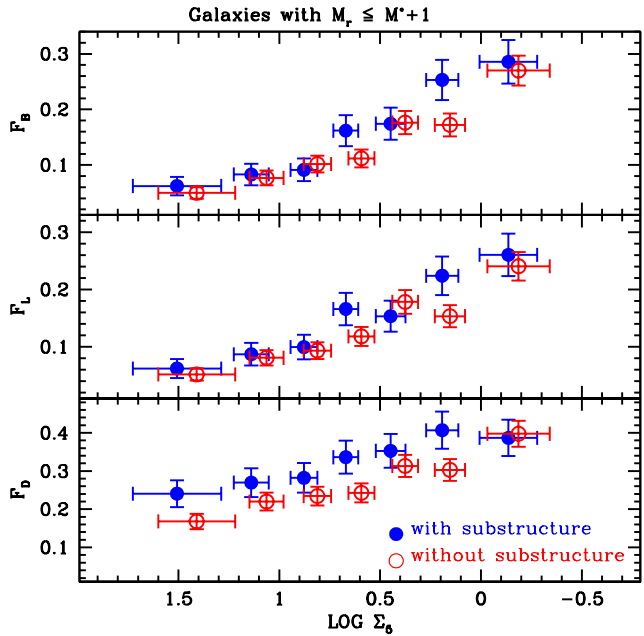}
\end{center}
\caption{Same as in Fig.~\ref{fig:denbins}, but splitting the sample in galaxies from
clusters with or without substructure.}
\label{fig:subbins}
\end{figure}

\begin{figure}
\begin{center}
\leavevmode
\includegraphics[width=3.5in]{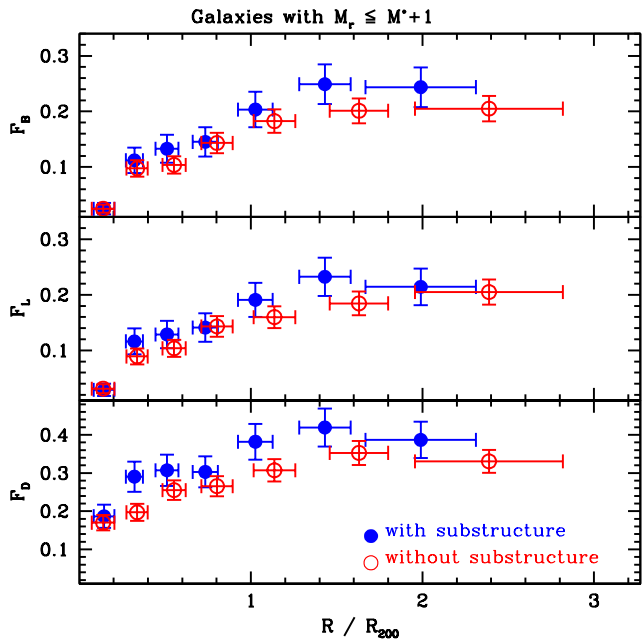}
\end{center}
\caption{Same as previous figure, but showing the variations with normalized distance 
to the cluster center ($R/R_{200}$).}
\label{fig:subbins2}
\end{figure}

In Figures ~\ref{fig:subbins} and ~\ref{fig:subbins2} we show the variations of 
$F_B, F_L$ and $F_D$ as a function of $\Sigma_5$ and $R/R_{200}$, as in the figures 
above. However, here we show the results for bright 
member galaxies coming from clusters 
with the presence or not of substructure (blue filled and red open circles, respectively).
Substructure is estimated using the $\Delta$ (or Dressler and Schectman) test 
(details in paper I).
It seems that in the most dense regions and within the virial radius there is no 
significant difference in the values of $F_B$ and $F_L$, but there might be 
more low concentration ($F_D$) galaxies in systems with substructure. For 
instance, at $R \sim 0.35 \times R_{200}$ $F_D \sim 29\%$ and $\sim 20\%$ for 
systems with and without substructure, respectively. For lower density regions
and outside $R_{200}$ we have slightly higher fractions ($F_B, F_L$ and 
$F_D$) for clusters with substructure. These results agree to the findings of
paper III, for which the systems are classified as Gaussian and non-Gaussian.

\subsection{Dependence on Cluster Mass}

\begin{figure}
\begin{center}
\leavevmode
\includegraphics[width=3.5in]{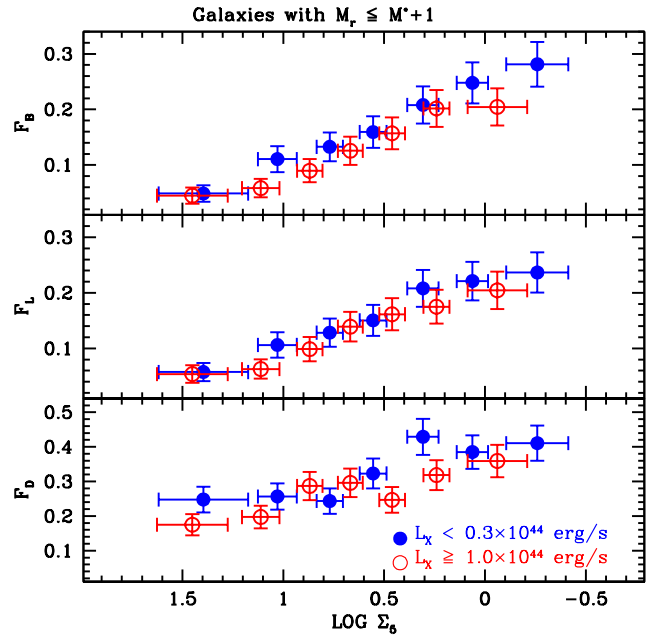}
\end{center}
\caption{Same as in Fig.~\ref{fig:denbins}, but splitting the sample in galaxies from
clusters with low or high gas content ($L_X < 0.3 \times 10^{44} erg s^{-1}$ or
$L_X \ge 1.0 \times 10^{44} erg s^{-1}$).}
\label{fig:lxbins}
\end{figure}

\begin{figure}
\begin{center}
\leavevmode
\includegraphics[width=3.5in]{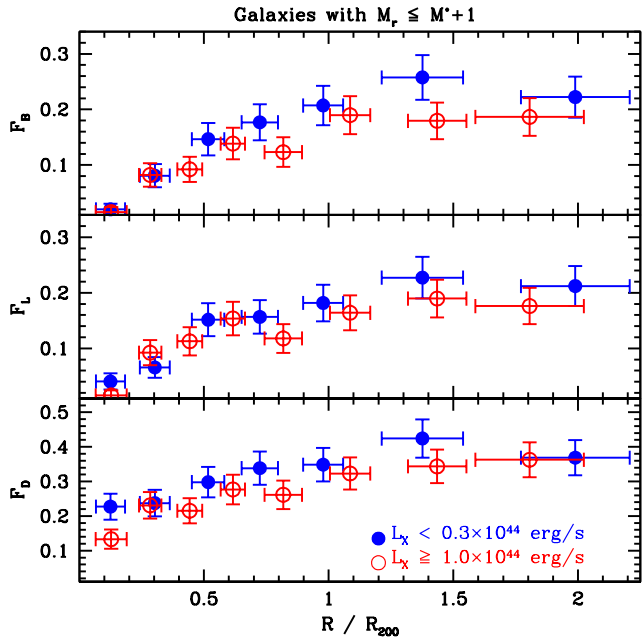}
\end{center}
\caption{Same as previous figure, but showing the variations with normalized distance 
to the cluster center ($R/R_{200}$).}
\label{fig:lxbins2}
\end{figure}

In this section we investigate the possible dependence of the galaxy population on 
the parent cluster mass, which we indicate by X-ray Luminosity and velocity dispersion.
On what follows we show the results obtained for systems at low and high mass ranges.
In Figures ~\ref{fig:lxbins} and ~\ref{fig:lxbins2} we show the variations of 
$F_B, F_L$ and $F_D$ as a function of $\Sigma_5$ and $R/R_{200}$. However, here we 
show the results for bright galaxies coming from clusters with low 
($L_X < 0.3 \times 10^{44} erg s^{-1}$) or high ($L_X \ge 1.0 \times 10^{44} erg s^{-1}$) 
X-ray luminosity (used as a mass indicator), blue filled and red open circles, 
respectively. From these plots we do not find any significant difference 
between the results for low or high $L_X$ systems. The exception is $F_D$ showing 
different results (although the results are noisier than for $F_B$ and $F_L$), 
but only in the central (within $R_{200}$) and most dense bins 
(especially the most central and dense). For instance, at 
$R \sim 0.1 \times R_{200}$ $F_D \sim 23\%$ and $\sim 13\%$ for systems with low and 
high X-ray luminosity, respectively. At $R \sim 0.7 \times R_{200}$ we find 
$F_D \sim 34\%$ and $\sim 28\%$, for low and high $L_X$, respectively. Note that $F_L$
is also slightly higher for low $L_X$ systems in the most central bin.

\begin{figure}
\begin{center}
\leavevmode
\includegraphics[width=3.5in]{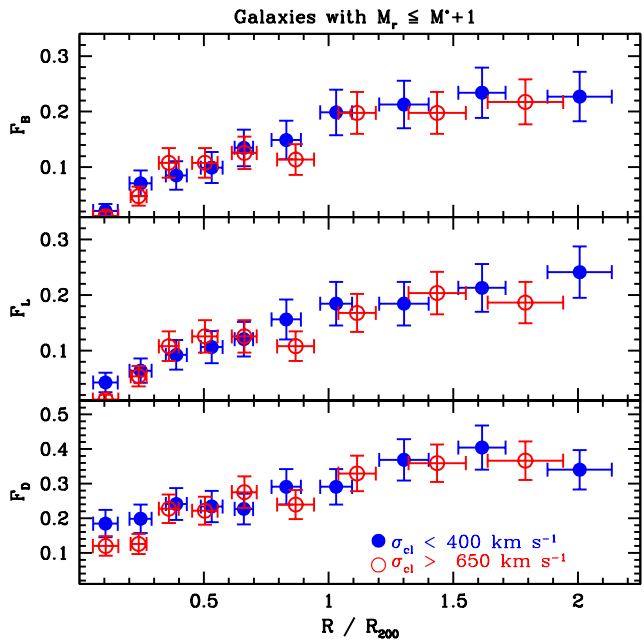}
\end{center}
\caption{Same as in Fig.~\ref{fig:radbins}, but splitting the sample in galaxies from
clusters with small or high mass, traced by the velocity dispersion. Blue 
filled circles are 
the results for groups ($<$ 400 km s$^{-1}$) and red empty circles for clusters ($>$ 650 km s$^{-1}$).}
\label{fig:vdispbins}
\end{figure}

In Figure~\ref{fig:vdispbins} we show again the variations of 
$F_B, F_L$ and $F_D$ as a function of $R/R_{200}$, but now we split the galaxies in
two mass ranges traced by velocity dispersion (having approximately the same number
of galaxies on each set). Namely groups with velocity
dispersion smaller than 400 km s$^{-1}$ (blue filled circles) and clusters with
$\sigma_{cl} >$ 650 km s$^{-1}$ (red open circles). According to this separation there are 84 low 
mass and 16 high mass systems from the original 152 in our sample. There is roughly
the same number of galaxies in the two samples, namely 1,685 galaxies in the groups and
1,497 in the high mass clusters. The median value for $R_{200}$ of the groups is 
0.99 Mpc, with a minimum of 0.51 Mpc and maximum value of 1.46 Mpc. The median mass
is 1.19 $ \times 10^{14} M_{\odot}$, with minimum and maximum values of 0.16 and 
3.84 $ \times 10^{14} M_{\odot}$, respectively. For the high mass sample the median, 
minimum and maximum values of $R_{200}$ are 2.00, 1.76 and 2.39 Mpc, respectively. In 
terms of mass the corresponding values are 9.82, 6.68 and 
16.93 $ \times 10^{14} M_{\odot}$.  There is no large difference in the
results for low and high mass systems, except for $F_D$ in the two most central bins
and $F_L$ in the central bin. At $R \la 0.3 \times R_{200}$ $F_D \sim 19\%$ for low
mass systems and $\sim 12\%$ for high mass objects. Hence, we find no large differences
regarding mass (using $L_X$ and $\sigma_{cl}$), except for $F_D$ and $F_L$ in the 
very central parts.

\subsection{Dependence on Galaxy Stellar Mass}

It is well known that galaxy stellar mass and environment are related, with massive
galaxies preferentially residing in the densest regions of the Universe. Many galaxy
properties also shown a correlation with stellar mass \citep{kau04}, so that it is 
important to try disentangling the connection between galaxy properties to 
environment and stellar mass. In order to do that we investigate in this section 
the environmental dependence of the galaxy population at fixed stellar mass. 
To obtain galaxy stellar mass estimates we used the stellar population synthesis
code STARLIGHT \citep{cid05}. This code fits an observed spectrum with a 
linear combination of a number of template spectra with known properties (see 
paper III for details).
As shown in $\S$4 most bright galaxies are massive, while most faint objects have 
small masses (the overlap for our sample occurs around $0.5 \times 10^{10} M_{\odot}$). 
Hence, the results shown next consider both bright and faint galaxies, with a 
combined sample having $M_r \le M^*+3$.

\begin{figure}
\begin{center}
\leavevmode
\includegraphics[width=3.5in]{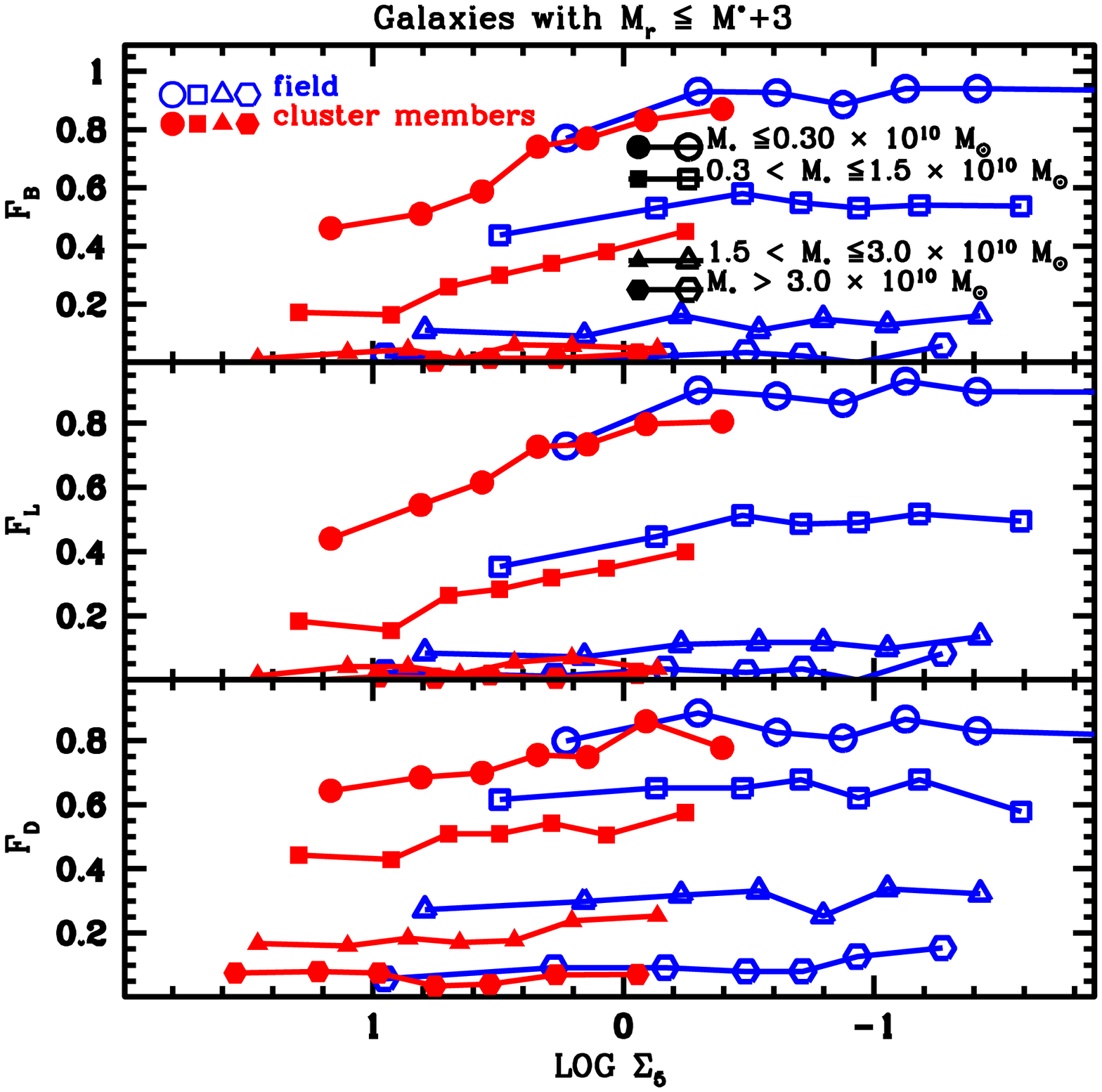}
\end{center}
\caption{Same as in Fig.~\ref{fig:denbins}, but splitting the sample according to the 
galaxy stellar mass. We consider four intervals: $M_* \le 0.3 \times 10^{10} M_{\odot}$ 
(circles), $0.3 < M_* \le 1.5 \times 10^{10} M_{\odot}$ (squares), 
$1.5 < M_* \le 3.0 \times 10^{10} M_{\odot}$ (triangles) and 
$M_* > 3.0 \times 10^{10} M_{\odot}$ (hexagons). As before, cluster members results are in
red (filled symbols) and field results in blue (open symbols). All galaxies with $M_r \le M^*+3$ (bright and faint) are
considered.}
\label{fig:den_stmassbins}
\end{figure}

\begin{figure}
\begin{center}
\leavevmode
\includegraphics[width=3.5in]{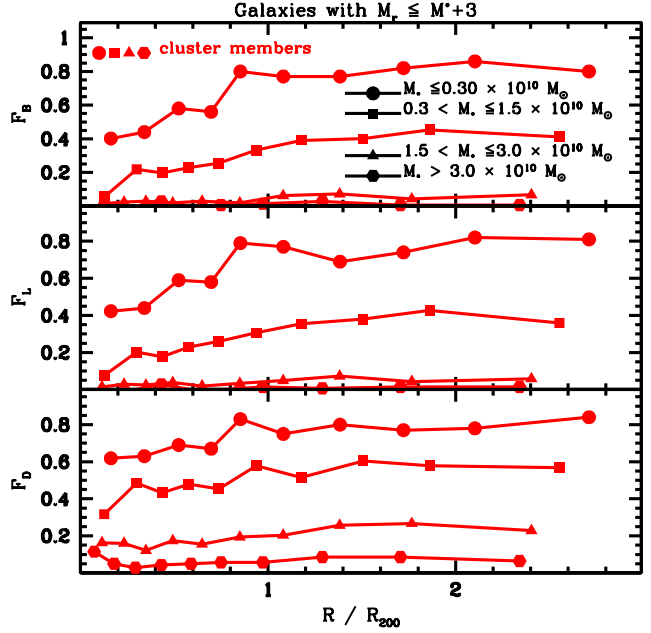}
\end{center}
\caption{Same as in Fig.~\ref{fig:radbins}, but splitting the sample according to the 
galaxy stellar mass. We consider four intervals: $M_* \le 0.3 \times 10^{10} M_{\odot}$ 
(circles), $0.3 < M_* \le 1.5 \times 10^{10} M_{\odot}$ (squares), 
$1.5 < M_* \le 3.0 \times 10^{10} M_{\odot}$ (triangles) and 
$M_* > 3.0 \times 10^{10} M_{\odot}$ (hexagons). All galaxies with $M_r \le M^*+3$ 
(bright and faint) are considered.}
\label{fig:rad_stmassbins}
\end{figure}

In Figures~\ref{fig:den_stmassbins} and~\ref{fig:rad_stmassbins} we show the 
variations of $F_B, F_L$ and $F_D$ as a function of $\Sigma_5$ and $R/R_{200}$. 
However, the results are now displayed for four different stellar mass ranges:
$M_* \le 0.3 \times 10^{10} M_{\odot}$ (circles), 
$0.3 < M_* \le 1.5 \times 10^{10} M_{\odot}$ (squares), 
$1.5 < M_* \le 3.0 \times 10^{10} M_{\odot}$ (triangles) and 
$M_* > 3.0 \times 10^{10} M_{\odot}$ (hexagons). Cluster members are show in
red (filled symbols), while field galaxies are in blue (open symbols). 

As expected, the largest fractions of blue, star-forming and discs are seen 
for the lower mass galaxies (the first two mass bins, 
$M_* \le 1.5 \times 10^{10} M_{\odot}$). The environmental variation for these 
objects is negligible for lower density regions, becoming relevant once galaxies
inhabit regions typical of cluster outskirts. Towards larger densities and cluster 
centers the variation in galaxy populations (for these low mass objects) becomes 
stronger (in agreement with \citealt{pre12}). However, 
the environmental dependence is steeper in terms of $F_B$ and 
$F_L$, being weaker (flatter) for $F_D$. On the contrary, the most massive galaxies 
have small values of $F_B, F_L$ and $F_D$ and also smaller variation with density 
and cluster radius. These results corroborate the idea that star formation is 
halted first in higher mass galaxies and that their environmental variation is 
small when compared to low mass galaxies. That is probably due to a combination 
of intrinsic galaxy evolution and the longer time these massive systems spend 
inside clusters.

\section{Relations of Physical Properties of Galaxies and their 
Environmental Variation}

There are well know relations between different photometric and spectral parameters 
of galaxies, which can be used to understand their formation and evolution. So,
as important as to see how those different parameters change with environment it is
to investigate the environmental dependence of these relations. That is the goal of
the current section, where we investigate relations involving spectral type, color,
stellar mass and size, for subsamples obtained relative to concentration, density,
cluster radial distance and mass.

\begin{figure*}
\begin{center}
\leavevmode
\includegraphics[width=7.0in]{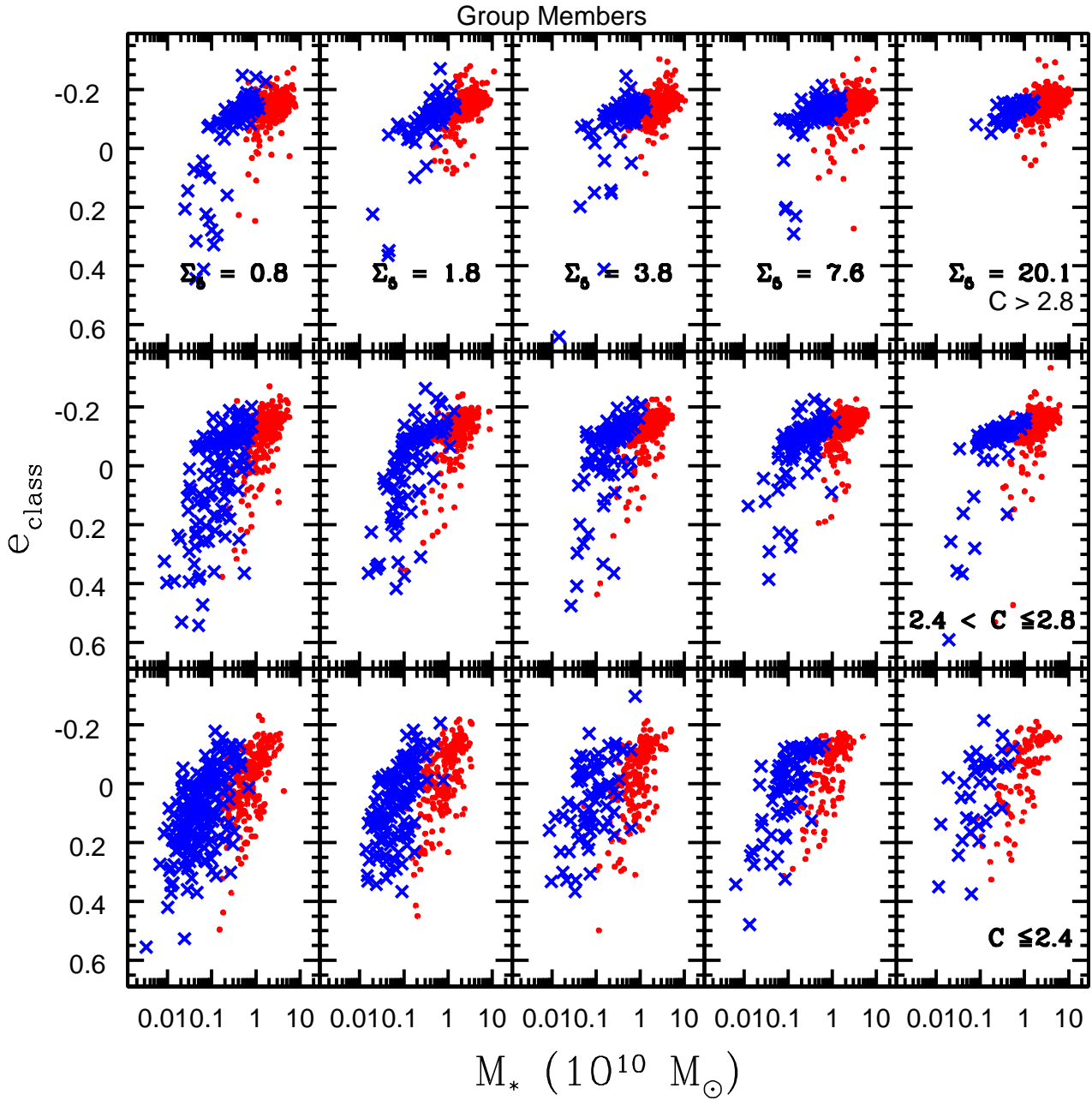}
\end{center}
\caption{Relation between $e_{class}$ and stellar mass in five local galaxy density
bins and for different ranges of concentration. Density increases from left to right 
and concentration from bottom to top panels. The $\Sigma_5$ bi-weight estimate of 
galaxies in each column is indicated 
in the top panels, ranging from $\Sigma_5 = 0.8$ to $\Sigma_5 = 20.1$ galaxies Mpc$^{-2}$. 
The lower panels have low concentration galaxies ($C \le 2.4$), the
middle panels have intermediate types ($2.4 < C \le 2.8$), while the top panels 
display bulges, with $C > 2.8$. Red small points indicate bright 
galaxies ($M_r \le M^*+1$) and blue large crosses are for faint objects 
($M^*+1 < M_r \le M^*+3$). The results consider only group members.}
\label{fig:ec_mstel}
\end{figure*}

In Fig.~\ref{fig:ec_mstel} we show the connection between the spectral type ($e_{class}$)
and stellar mass in five local galaxy density bins and for different ranges of 
concentration. Density increases left-right and concentration from bottom-top. The 
$\Sigma_5$ bi-weight estimate of galaxies in each column is indicated 
in the top panels, ranging from $\Sigma_5 = 0.8$ to $\Sigma_5 = 20.1$ galaxies Mpc$^{-2}$. 
In the lower panels we show low concentration galaxies ($C \le 2.4$), 
while intermediate types ($2.4 < C \le 2.8$) are displayed in the middle panels and
bulges ($C > 2.8$) are shown in the top panels. Red small points indicate bright 
galaxies ($M_r \le M^*+1$) and blue large crosses are for faint objects 
($M^*+1 < M_r \le M^*+3$). The results consider only group members. In $\S$2.3 we
call low concentration galaxies ($C \le 2.4$) as {\it classical discs} and the high
concentration objects as {\it classical bulges}. From now on, for simplicity, we
are calling these two populations just as discs and bulges, respectively. 

From Fig.~\ref{fig:ec_mstel} we can see that bulge-dominated objects ($C > 2.8$) are 
mostly passive objects ($e_{class} < -0.05$). The exception is for a few star-forming 
galaxies found mainly in the lowest density bin. Those objects are mainly faint and 
with a small stellar mass content. However, a few bright star-forming objects (red 
points) are also found at higher local densities. For disc-dominated objects 
($C \le 2.4$) most points are consistent with star-forming low mass galaxies 
($e_{class} > -0.05$). We can also see that as density increases the median stellar
mass too. The most dramatic environmental variation is seen for intermediate type 
galaxies, which become more massive and, most important, passive as we go from
the left to right panels.

\begin{figure*}
\begin{center}
\leavevmode
\includegraphics[width=7.0in]{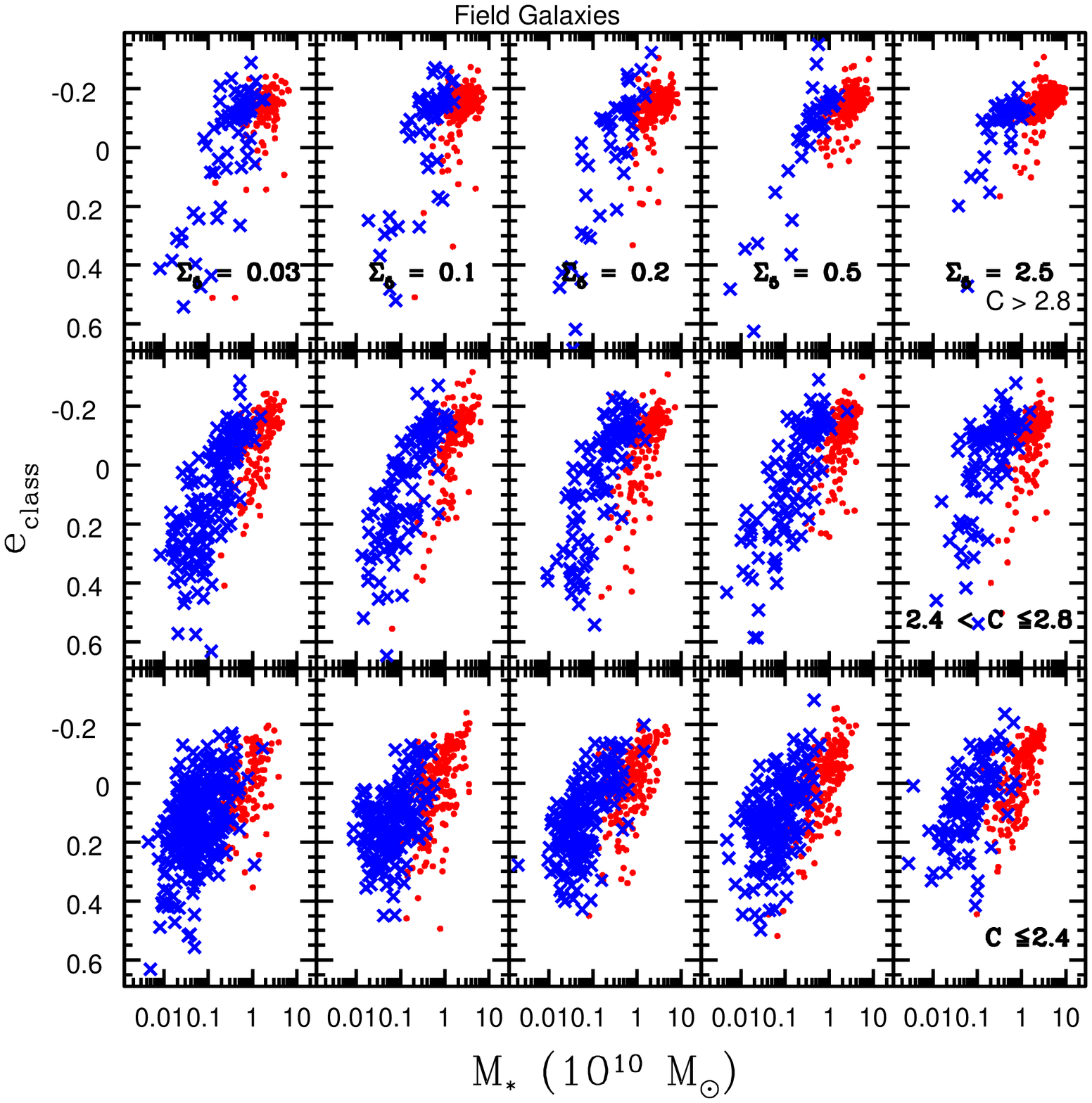}
\end{center}
\caption{Same as previous figure, but showing the results for field galaxies.}
\label{fig:ec_mstel2}
\end{figure*}

Investigating field galaxies (figure~\ref{fig:ec_mstel2}) we first notice the five
density ranges are much different, ranging from $\Sigma_5 = 0.03$ to $\Sigma_5 = 2.5$ 
galaxies Mpc$^{-2}$. Hence, the  highest density bin 
is comparable to the second poorest for clusters (typical of their outskirts). 
Nonetheless, we can see the behavior for galaxies at similar densities (in 
groups or field, in figures~\ref{fig:ec_mstel} and \ref{fig:ec_mstel2}) show 
similar relations. For field galaxies we notice that star-forming objects are 
dominant even for intermediate types ($2.4 < C \le 2.8$) and there is a 
steeper variation of $e_{class}$ as a function of $M_*$ for intermediate types in the 
three lowest density bins.

\begin{figure*}
\begin{center}
\leavevmode
\includegraphics[width=7.0in]{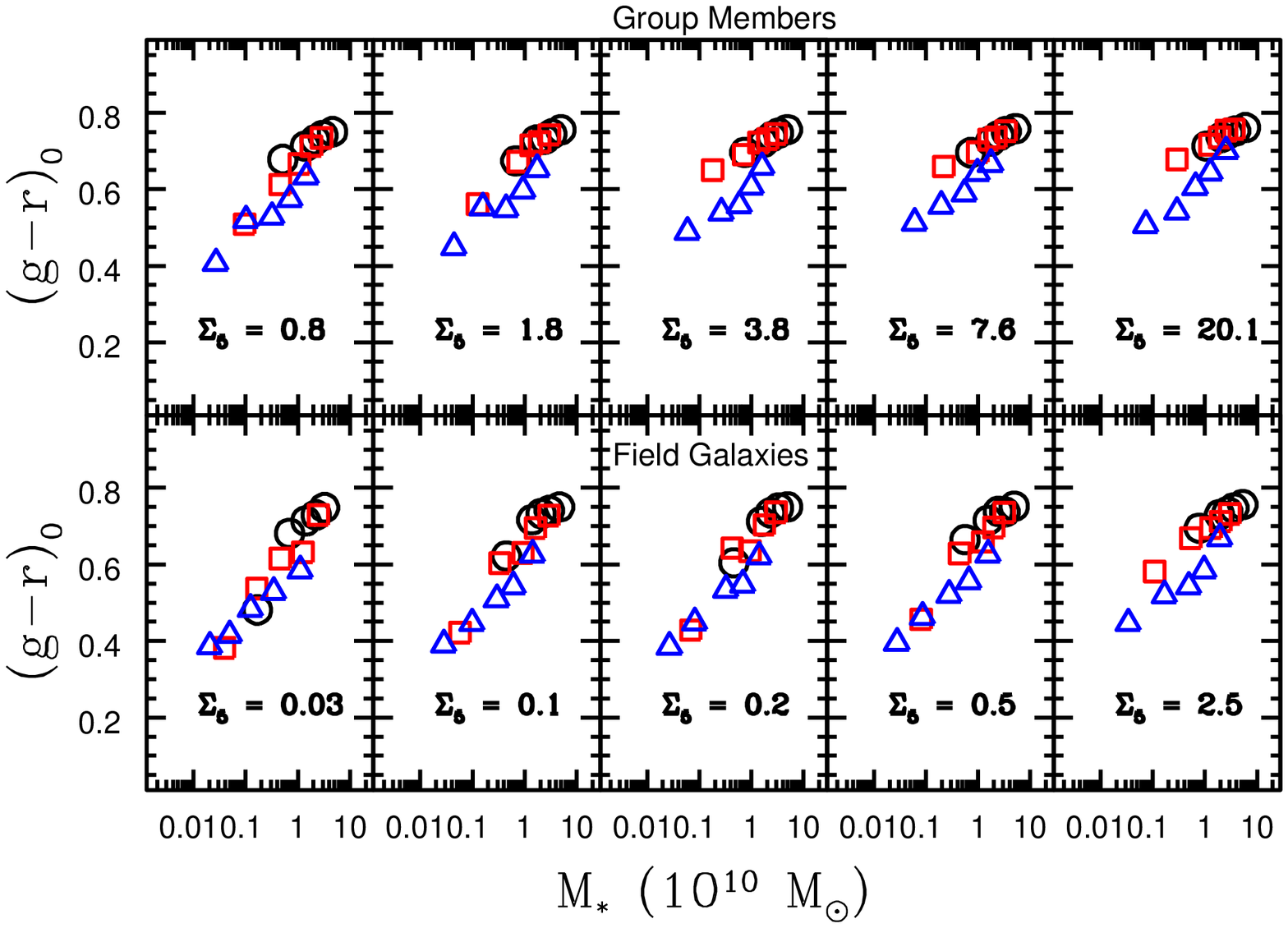}
\end{center}
\caption{Relation between rest-frame color $(g-r)_0$ and stellar mass in five 
local galaxy density bins and for different ranges of concentration. Density increases 
from left to right. Differently from previous two figures we only display the
binned results (considering stellar mass) on each panel. Low concentration 
galaxies ($C \le 2.4$) are displayed as blue triangles, intermediate types 
($2.4 < C \le 2.8$) as red squares and bulges ($C > 2.8$) as dark circles. 
The top panels show the results for group members and the bottom panels are 
for field galaxies.}
\label{fig:gr_mstel}
\end{figure*}

Figure~\ref{fig:gr_mstel} is similar to the previous two, but we show the relation 
between the rest-frame color $(g-r)_0$ (a rough star formation 
indicator) and stellar mass in five local galaxy density 
bins and for different ranges of concentration. As before, density increases from
left to right. Differently from previous two figures we only display the binned 
results (considering stellar mass) on each panel. We use different symbols for
the three ranges of concentration. Blue triangles are for low concentration galaxies 
($C \le 2.4$), red squares represent intermediate types ($2.4 < C \le 2.8$) and dark
circles display bulges ($C > 2.8$). In the top panels we have the results for group
members and in the bottom for field galaxies. The most striking variations are seen 
for intermediate concentration objects ($2.4 < C \le 2.8$), both for field and group 
galaxies. As density increases (left to right panels) these objects (shown by red 
squares) become progressively more massive and most important red. We can see the low
mass intermediate type galaxies assume redder colors as density increases. That is true 
even for the field galaxies, which are restricted to lower density environments. We can
see a similar behavior for low mass discs (blue triangles, $C \le 2.4$), but the effect 
is not as strong as for the red squares, especially for field galaxies. Bulges (dark 
circles, $C > 2.8$) are mostly massive and red objects. For groups that is true for all
environments, even in the outskirts (low densities), while we see the lower mass field 
bulges are bluer in the low density environments (first three density bins).

\begin{figure*}
\begin{center}
\leavevmode
\includegraphics[width=7.0in]{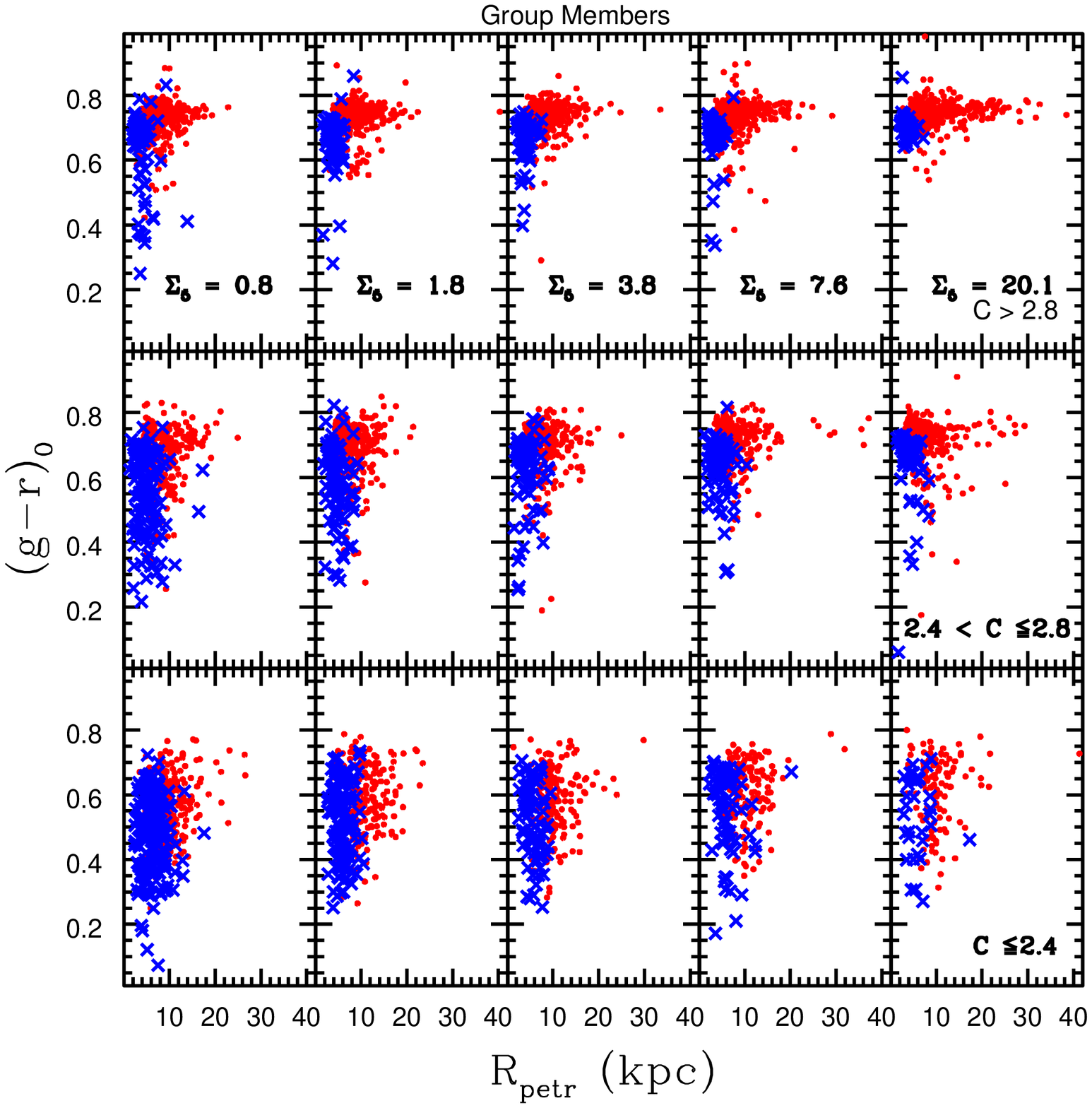}
\end{center}
\caption{Relation between rest-frame color $(g-r)_0$ and petrosian radius in five 
local galaxy density bins and for different ranges of concentration. Density increases 
from left to right and concentration from bottom to top panels. The $\Sigma_5$ 
bi-weight estimate of galaxies in each column is indicated 
in the top panels, ranging from $\Sigma_5 = 0.8$ to $\Sigma_5 = 20.1$ galaxies Mpc$^{-2}$. 
The lower panels have low concentration galaxies ($C \le 2.4$), the
middle panels have intermediate types ($2.4 < C \le 2.8$), while the top panels 
display bulges, with $C > 2.8$. Red small points indicate bright 
galaxies ($M_r \le M^*+1$) and blue large crosses are for faint objects 
($M^*+1 < M_r \le M^*+3$). The results consider only group members.}
\label{fig:gr_rpet}
\end{figure*}

\begin{figure*}
\begin{center}
\leavevmode
\includegraphics[width=7.0in]{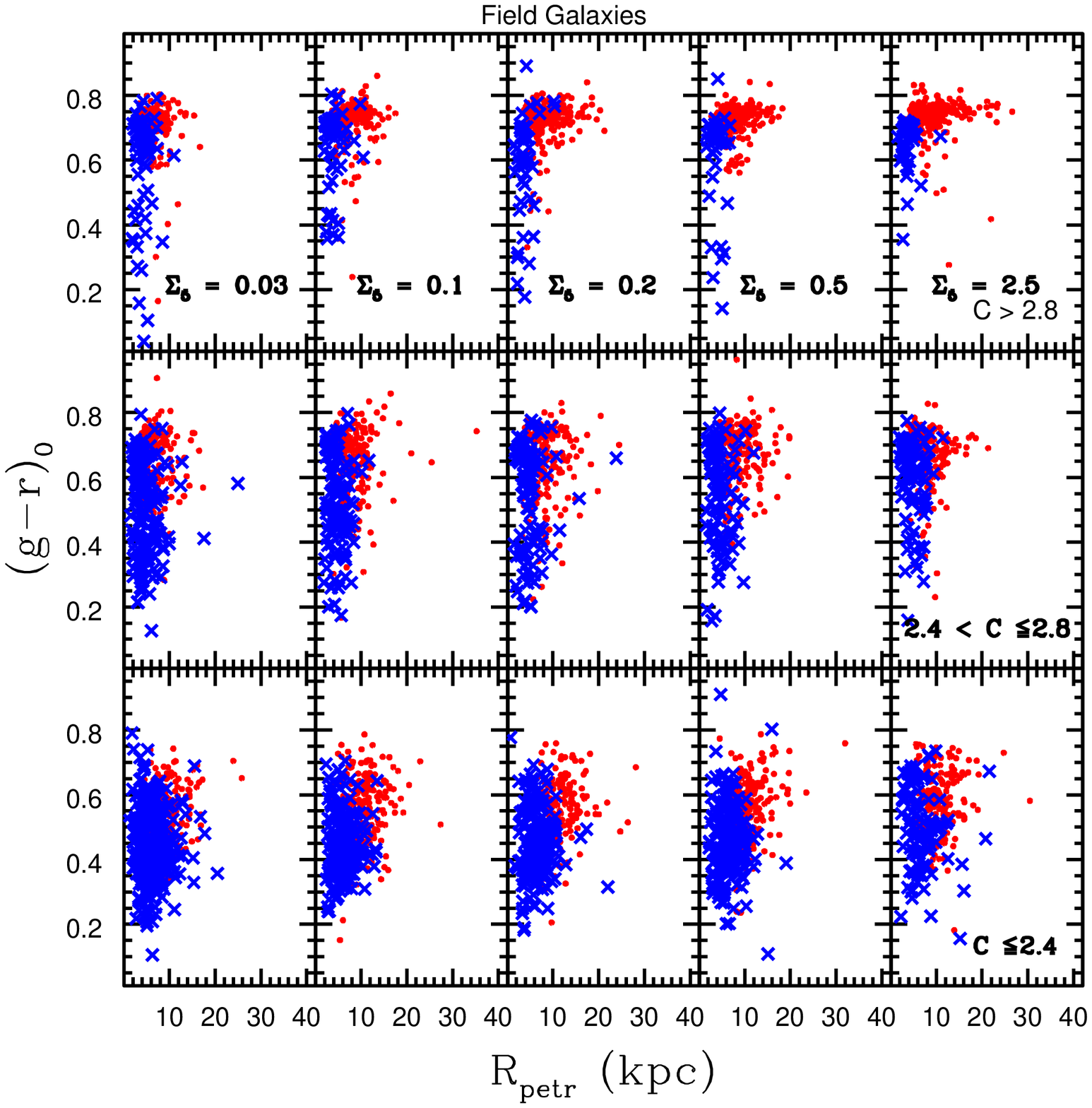}
\end{center}
\caption{Same as previous figure, but showing the results for field galaxies.}
\label{fig:gr_rpet2}
\end{figure*}

We also investigate how the rest-frame color $(g-r)_0$ correlates with galaxy size 
(indicated by the petrosian radius in the $r-$band). That is shown for group 
members in Fig.~\ref{fig:gr_rpet} in the same five density bins and for the three 
concentration ranges as above. As in Fig.~\ref{fig:ec_mstel}
red small points are for bright galaxies and blue large crosses for faint. Fig.~\ref{fig:gr_rpet2} is analogous
to Fig.~\ref{fig:gr_rpet}, but showing the results for field galaxies.

From the group members, as we go bottom-up and left-right we 
can see galaxies forming a tight relation in the
color-size space, with large and more concentrated galaxies becoming redder
(see \citealt{cyp06}). 
In other words, as galaxies get to denser environments and grow in size, 
their star formation decreases. Besides that we see that the fraction of 
blue galaxies is also larger for the middle and bottom panels. If 
we consider $(g-r)_0 = 0.75$ as a typical color for red 
galaxies \citep{lop07, fuk95} and call blue galaxies with $(g-r)_0 \la 0.65$ (typical 
color minus 0.10) we see there is only a small number of blue objects that are also 
bulges in the three most dense bins (upper right panels). In all the other panels 
the number of blue objects increases, even for bulges in the two lowest density bins 
(upper left panels). With a few exceptions, objects with $R_{petr} \ga 20$ kpc are 
only seen in the two most dense bins, for intermediate types and bulges. An 
interest exception is a large disc ($R_{petr} \ga 40$ kpc) seen in the lower right panel.

Considering the field galaxies (Fig.~\ref{fig:gr_rpet2}) we see the color-radius
sequence is starting to form for bulges in high densities (upper right), but still
with many blue galaxies and with a small proportion of large objects 
($R_{petr} \ga 20$ kpc). In all the remaining panels we see a large fraction of
blue galaxies and nearly no galaxies with $R_{petr} \ga 20$ kpc.

\begin{figure*}
\begin{center}
\leavevmode
\includegraphics[width=7.0in]{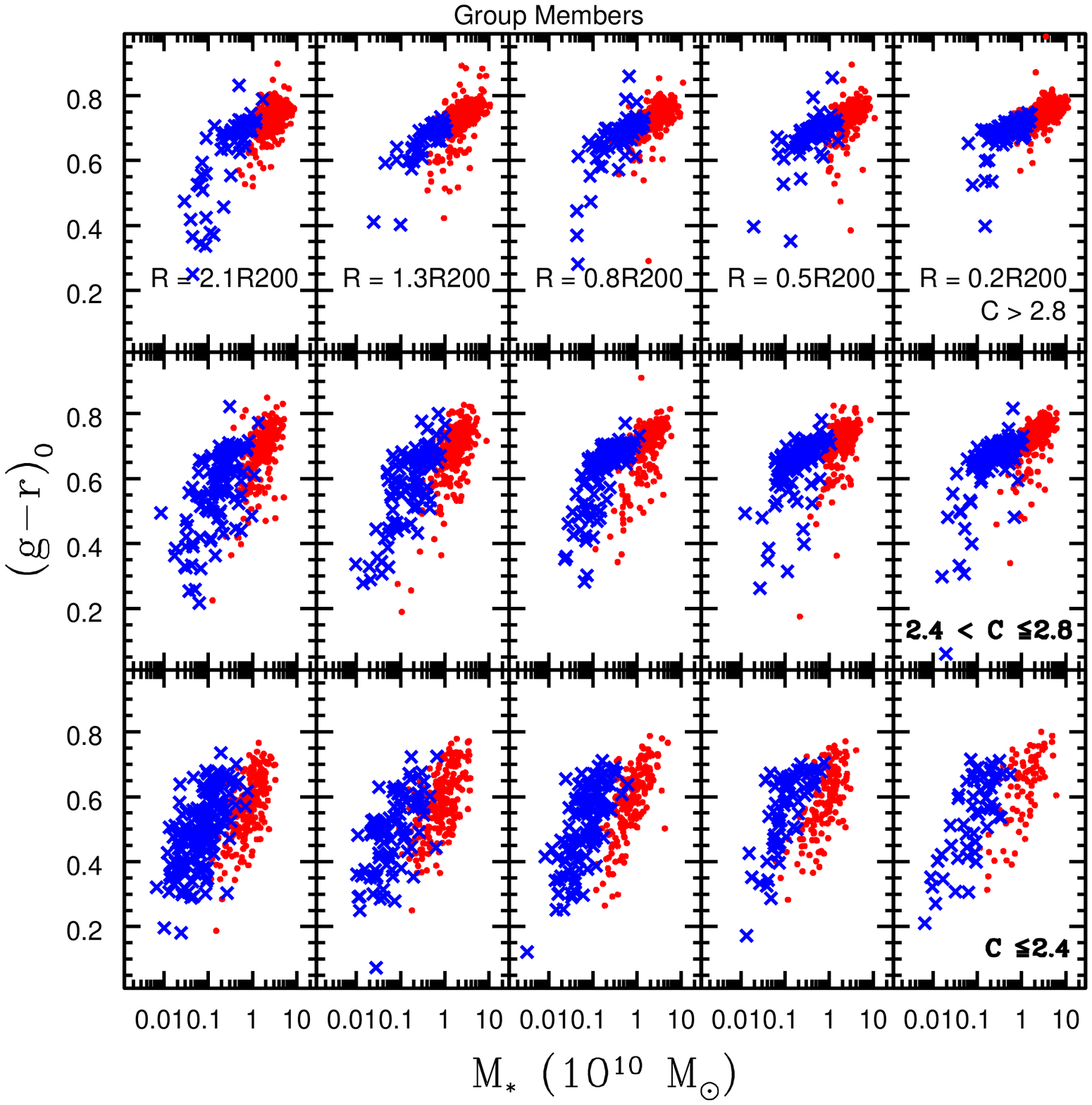}
\end{center}
\caption{Relation between $(g-r)_0$ and stellar mass in five bins of radial 
separation from the cluster center (in units of $R_{200}$) and for different ranges 
of concentration. Radial separation decreases from left to right 
and concentration increases from bottom to top panels. The radial offset bi-weight 
estimate of galaxies in each column is indicated in the top panels, ranging from 
$R = 2.1 R_{200}$ to $R = 0.2 R_{200}$. The lower panels have low concentration 
galaxies ($C \le 2.4$), the middle panels have intermediate types ($2.4 < C \le 2.8$), 
while the top panels display bulges, with $C > 2.8$. Red small points indicate bright 
galaxies ($M_r \le M^*+1$) and blue large crosses are for faint objects 
($M^*+1 < M_r \le M^*+3$). The results consider only group members.}
\label{fig:gr_mstel2}
\end{figure*}

Fig.~\ref{fig:gr_mstel2} is similar to the ones right above, but now we separate
galaxies according to the distance to the center of their parent group. We show the 
relation between $(g-r)_0$ and $M_*$, also split according to $C$  and for bright 
(red) and faint (blue) objects. The red sequence is easily seen for bulges (upper 
panels) and it may be seen for intermediate types in the central parts of groups 
(middle right panels). It is interesting to see that for bulges in the outskirts 
(upper left) the red sequence has a much larger scatter and there is a non-negligible 
fraction of blue bulges. In the outskirts the discs (lower left) are mainly blue. 
In general, the results of this figure corroborate what was seen in 
Fig.~\ref{fig:gr_mstel}, splitting galaxies according to local density.

\subsection{Possible Variation With Cluster Mass}

\begin{figure*}
\begin{center}
\leavevmode
\includegraphics[width=7.0in]{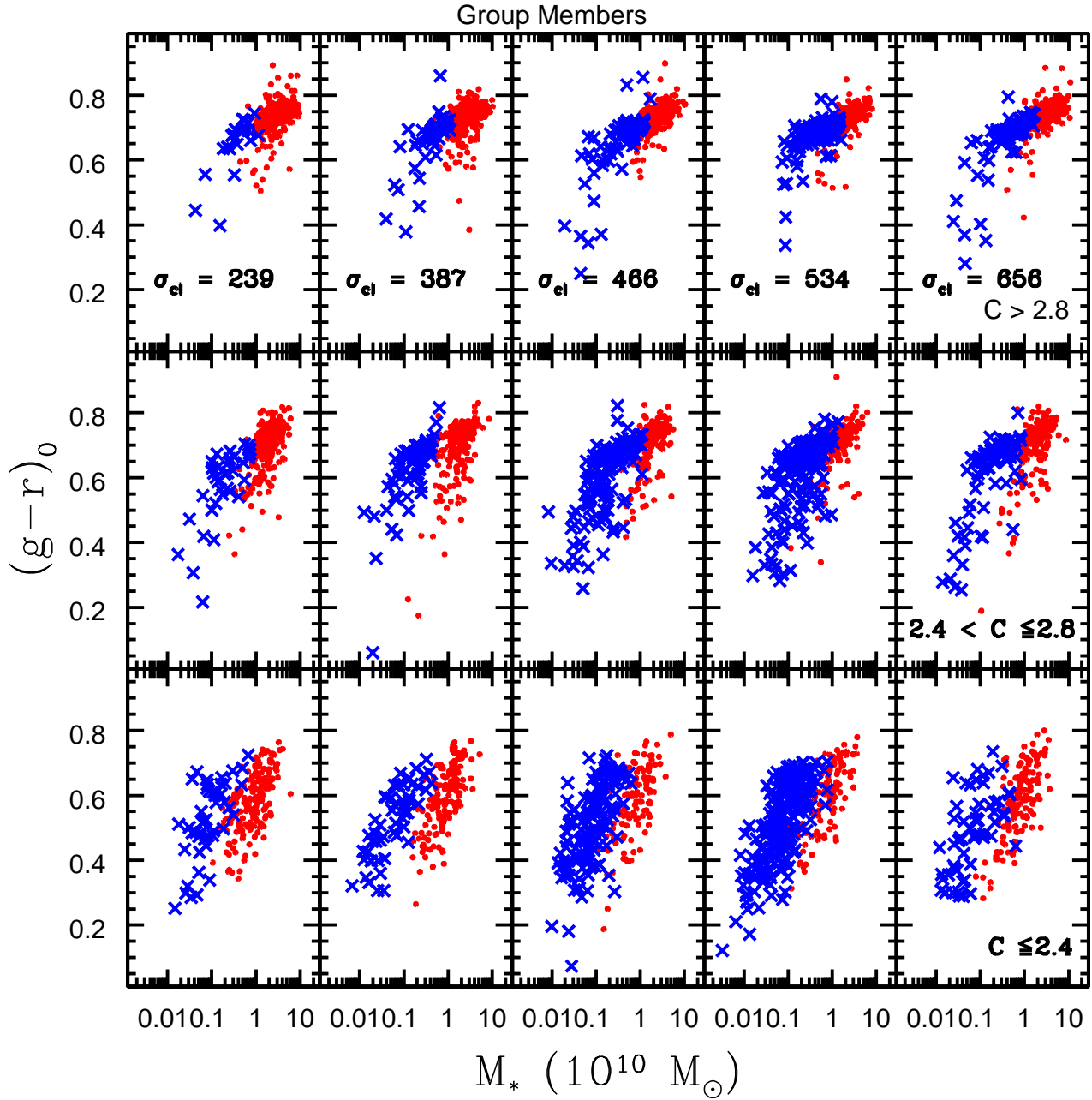}
\end{center}
\caption{Relation between $(g-r)_0$ and stellar mass in five bins of cluster velocity 
dispersion (in units of km s$^{-1}$) and for different ranges 
of concentration. Velocity dispersion ($\sigma_{cl}$) increases from left to right 
and concentration from bottom to top panels. The $\sigma_{cl}$ bi-weight 
estimate in each column is indicated in the top panels, ranging from 
$\sigma_{cl} = 239$ to $\sigma_{cl} = 656$ km s$^{-1}$. The lower panels have low 
concentration galaxies ($C \le 2.4$), the middle panels have intermediate types 
($2.4 < C \le 2.8$), while the top panels display bulges, with $C > 2.8$. Red small 
points indicate bright galaxies ($M_r \le M^*+1$) and blue large crosses are for 
faint objects ($M^*+1 < M_r \le M^*+3$). The results consider only group members.}
\label{fig:ec_magrvd}
\end{figure*}

Besides the dependence on environment traced by local galaxy density or 
radial distance (to the parent cluster) we have also investigated if the 
relations between physical properties of galaxies depend on the parent 
cluster mass, traced by velocity dispersion ($\sigma_{cl}$). Fig.~\ref{fig:ec_magrvd} 
is similar to Fig.~\ref{fig:gr_mstel}, but now galaxies are separated accordingly to 
the velocity dispersion of the parent group/cluster. We consider five bins of 
($\sigma_{cl}$). The $\sigma_{cl}$ bi-weight estimate in each column is indicated 
in the top panels, ranging from $\sigma_{cl} = 239$ to $\sigma_{cl} = 656$ km s$^{-1}$. 
The lower panels have low concentration galaxies ($C \le 2.4$), the middle panels 
have intermediate types ($2.4 < C \le 2.8$), while the top panels display bulges, 
with $C > 2.8$. As before, red small points indicate bright galaxies ($M_r \le M^*+1$) 
and blue large crosses are for faint objects ($M^*+1 < M_r \le M^*+3$). The results consider 
only group members. From this figure, the main difference regarding $\sigma_{cl}$
may only be the scatter of the red sequence, which might be smaller for higher
mass clusters. More massive clusters may also have more faint low mass bulges (blue
crosses), as seen in the upper panels. Considering galaxies to first be 
pre-processed in groups, their star formation and color transformation may be 
further reduced when their parent groups are accreted into larger clusters. 
Those results would contradict
the ones from \citet{val11}. They find no dependence of 
several different properties of the cluster RS (including the scatter) 
with parent halo mass. However, the comparison is not straightforward as their
sample is restricted to clusters with $\sigma_P > 500$ km s$^{-1}$ and for
galaxies within $0.5 \times R_{200}$.

From this plot we conclude that there is no large difference (expect for the scatter 
in the red sequence and large number of low mass bulges) in the relations of 
galaxy properties for groups and
massive clusters, corroborating the idea that galaxies are pre-processed in groups
before those are assembled in larger systems. But no large modification occurs after
that, except for the tightening of the relations involving luminosity (color-magnitude,
for instance). That may be due to a gradual decrease in the star formation rate
of galaxies as they reside in larger structures.

\section{Summary \& Discussion}

In this work we investigate the impact of environment on the abundance of galaxy 
populations and on the relations of their physical properties (such as color, 
stellar mass, size and spectral classification). Our sample consists of 6,415 
cluster members split in
two luminosity and redshift ranges, being 5,106 bright galaxies ($M_r \le M^*+1$;
at $z \le 0.100$) and 1,309 faint galaxies ($M^*+1 < M_r \le M^*+3$; at $z \le 0.045$).
These galaxies are selected from a total of 152 clusters in the local Universe with 
velocity dispersion in the range $150 \la \sigma_P \la 950$ km s$^{-1}$ (in terms of 
mass being $10^{13} \la M_{200} \la 10^{15} M_{\odot}$). The local environment is traced
by $\Sigma_5$ (a nearest neighbour method), while the global environment is related to
cluster parameters, such as distance to the cluster center, substructure, X-ray 
luminosity and velocity dispersion. The analysis regarding the local environment 
also shows density estimates ($\Sigma_5$) of field galaxies. Besides the influence of 
the environment we investigated the dependence of galaxy populations relative to
galaxy stellar mass. Our main results are:

\begin{enumerate}

\item The fractions of blue ($F_B$), high $e_{class}$ (or star-forming, $F_L$), and 
low concentration (or disc-dominated, $F_D$) galaxies show a strong variation with 
local galaxy density, from cluster cores to the field. For bright cluster 
members $F_B$ ranges from 
$\sim 5\%$ to $\sim 27\%$, while for field galaxies the variation is from $\sim 20\%$ 
to $\sim 44\%$. The results are similar for $F_L$, but the typical values are 
higher for $F_D$ (ranges from $\sim 20\%$ to $\sim 40\%$ for bright member 
galaxies and from $\sim 33\%$ to $\sim 53\%$ for field galaxies). There is a 
smooth transition between cluster and field galaxies, and at similar densities 
the fractions are higher for field galaxies.

\item For faint objects there is a strong variation with density in 
$F_B$, $F_L$ and $F_D$ for cluster members. However, for field galaxies these 
fractions are nearly constant. $F_B$ and $F_L$ range from $\ga 20\%$ to $\la 70\%$ 
for cluster members, increasing at most to $80\%$ for field objects. $F_D$ changes 
from $\sim 40\%$ to $\la 70\%$ for cluster members, being that the field value. 
Note that for all the parameters there is a minimum fraction ($\sim 20\%$ for
$F_B$ and $F_L$ and $\sim 40\%$ for $F_D$) in the most dense regions, indicating that 
the formation of dwarfs in the central parts of clusters might be effective.

\item For bright and faint galaxies $F_B$ and $F_L$ are similar and generally
lower than $F_D$, especially for cluster objects. That corroborate other results
in the literature indicating that color and star-formation are affected
faster than structural properties \citep{hom06, bun10}.

\item Considering the distance to the cluster center, for bright galaxies, the
populations do not change significantly until they fall within the virial radius.
Another abrupt variation seems to happen at $0.3 \times R_{200}$ (in agreement with
\citealt{got03}). Within that radius the
values of $F_B$, $F_L$ and $F_D$ decrease faster. We also find the clusters' cores 
to be nearly devoided of bright, blue and star-forming objects ($F_B = F_L \la 2\%$).

\item For faint objects a similar behavior is seen, but with different regimes.
From large to small radius $F_B, F_L$ and $F_D$ are approximately constant
until $\sim 1.8 \times R_{200}$, then decreasing, but remaining roughly constant
between $\sim 1.8 \times R_{200}$ and $\sim R_{200}$. Then there is a steep decline 
until $\sim 0.4 \times R_{200}$. Within that radius there is a slight 
rise in the values of $F_B, F_L$ and $F_D$ before reaching the minimum value in 
the most central bin. That rise might be associated to
the formation of small galaxies in the core of clusters, from tidal debris 
or due to ram pressure stripping of larger galaxies \citep{mas05, rib13}.

\item As a function of radius $F_D$ is also generally higher than $F_B$ and $F_L$,
corroborating the slower change in structural parameters. We also show that the 
dependence with radius goes to large radius ($\ga 3 \times R_{200}$), although
the variation is much stronger within the virial radius. That agrees with 
\citet{bah13}, suggesting that some processes like {\em overshooting}
and ram pressure may be relevant in the outskirts.

\item Separating the clusters according to the presence of substructure we find 
that within the virial radius $F_D$ seems to be higher for systems with 
substructure. Outside $R_{200}$ and for lower density regions there are slightly 
higher fractions for objects with substructure.

\item Investigating the dependence to cluster mass, traced by $L_X$ or
velocity dispersion we find no significant difference between the results for 
groups and massive clusters. The exceptions are $F_D$ and $F_L$ in the most
central bins, which agrees with other results in the literature ({\it {e.g.}} 
\citealt{cal12}).

\item Regarding galaxy stellar mass we find the largest fractions of blue, 
star-forming and discs are seen for the lower mass galaxies. The environmental 
variation for these objects is negligible for lower density regions, becoming 
relevant once galaxies reach regions typical of cluster outskirts, increasing 
further to higher densities. The variation of $F_D$ is not as steep as for
$F_B$ and $F_L$.

\item On the other hand, the most massive galaxies have small 
values of $F_B, F_L$ and $F_D$ and also smaller variation with density and 
cluster radius. These results corroborate the idea that star formation is halted
first in higher mass galaxies and that their environmental variation is small when
compared to low mass galaxies. That is probably due to a combination of intrinsic
galaxy evolution and the longer time these massive systems spend inside clusters.

\item The relation between $e_{class}$ and stellar mass shows a tight 
connection for bulges, both at 
low and high densities. For discs the relation may be forming at high densities for
cluster members. However, the most dramatic environmental variation is seen for 
intermediate type galaxies. They become more massive and, most important, passive 
as density increases. We also find that for the most dense regions (either in 
clusters or field) there is a non-negligible population of bulges that are 
classified as star-forming (those are mainly faint and low-mass galaxies). For 
field objects the fractions of star-forming are higher and the connection between 
$e_{class}$ and $M_*$ is steeper than for cluster galaxies.

\item The connection between rest-frame color $(g-r)_0$ and stellar mass 
also indicates the largest
environmental variation is seen for intermediate type objects ($2.4 < C \le 2.8$)
when compared to classical discs ($C \le 2.4$) and bulges ($C > 2.8$). Galaxies
called intermediate type progressively become more massive and especially redder
as density increases. That is true even for the field galaxies (restricted to 
lower density environments). 

\item The color-size relation also shows a tight connection as concentration and
density increases. However, large galaxies ($R_{petr} \ga 20$ kpc) are generally seen 
for intermediate types and cluster bulges in the two highest density bins. The 
fraction of blue galaxies is also larger for systems in low-density 
regions or discs in high densities (both for field and cluster galaxies).

\item Separating the color-mass (galaxy stellar mass) relation according to the 
mass of the parent system we find no significant difference between the results 
for group or high mass cluster galaxies. The only exception being the scatter of the 
color-mass relation and an excess of low mass bulges in massive clusters. Those 
results also indicate that pre-processing in groups should be very effective.

\end{enumerate}

A controversial issue in the literature regards the variation of
galaxy populations with parent halo mass (see \citealt{del12, val11, cal12, hoy12}). 
Although
some works find no variation with halo mass, others indicate a significant dependence. 
Part of this discrepancy may be due to different 
mass and radial ranges sampled by each study. In the current work we have objects from
small groups to large clusters ($10^{13} \la M_{200} \la 10^{15} M_{\odot}$), and spanning 
radial distances up to $\ga 3 \times R_{200}$. We find that only in 
the very central parts
there is a small difference in $F_D$ and $F_L$ for low and high mass clusters.
That agrees with \citet{cal12}, but slightly disagrees with \citet{val11} and 
\citet{hoy12} (although they
find a small dependence of the spiral fraction on halo mass). For the first case the
mass range sampled is very different (\citealt{val11} only have large
clusters), while in the second the morphological indicator (as well as the member 
selection) used is different than ours. However, all these results reinforce the 
idea that galaxy transformation happens predominantly on group scales, with large 
clusters being the result of an hierarchical aggregation of smaller systems 
(with no significant further morphological transformation, \citealt{hoy12}). 

The results from \citet{lac13} are 
complementary to this idea, indicating that star 
formation is quenched in the group scale, but morphological transformation
is a separate process, occurring in clusters. In other words, environmental 
galaxy transformations can be divided in two steps. First, star formation is
halted in discs residing in relatively low density environments. Second, a 
morphological transformation from disc to bulge dominated systems occur
at higher densities. The authors consider groups as low density and high mass
clusters as high density environments. But groups can be as dense
as clusters (especially compact groups, CG). For instance, \citet{coe12}
consider that pre-processing is even more effective in compact groups compared to 
loose groups (LG). That could be due to the high densities and low velocity 
dispersion typical of CGs. Note that \citet{lac13} are interested 
on classical bulge plus disc galaxies (sample with 12500 objects).

Our results add to the above conclusions, as we find no significant differences 
between the values of $F_B$, $F_L$ and $F_D$ for groups and clusters (except for 
$F_D$ in the most central regions). We also find that $F_D$ is larger than $F_B$ and 
$F_L$, for bright and faint galaxies. Hence, in agreement to 
\citet{val11}, our results point to a scenario on which
local density is the main driver for galaxy evolution and not the parent halo mass.
This evolution happens as pre-processing in the group scale with star formation 
quenching and, to a lesser extent, morphological transformation. However, 
the second effect takes longer for being effective ($F_D > F_B,F_L$) and shows a 
small segregation between groups and clusters in their central regions (most 
dense in the Universe); as $F_D$ is slightly higher for groups than for high 
mass clusters. That can be seen from Figures~\ref{fig:lxbins},~\ref{fig:lxbins2} 
and~\ref{fig:vdispbins} (so, both as a function of density and radius). What we see 
in the current work does not contradict \citet{lac13} or 
\citet{val11}. The former consider a sample of classical bulge 
plus disc galaxies only, while we have galaxies of all morphologies. 
\citet{val11} work with high mass clusters, consider the most 
central region ($0.5 \times R_{200}$) only and are restricted to red sequence 
galaxies, while we consider a broader mass range and a larger region around 
clusters.

The pre-processing in groups can be explained by different mechanisms, such as mergers,
strangulation and ram pressure, that can accelerate star formation, remove the gas 
reservoir and destroy galactic discs \citep{kau04}. Internal processes 
associated to the stellar mass are also relevant to halt star formation. Within 
larger systems galaxy mergers are less probable, but galaxies may still be 
transformed due to the cumulative effects of successive weak encounters, tidal 
stripping by the cluster gravitational potential and interaction with the denser 
ICM.

The smooth transition between cluster and field, seen in Figures~\ref{fig:denbins} 
and~\ref{fig:denbins2}, indicate that although cluster edges are hard to define,
density is indeed a key parameter for galaxy transformation. For bright galaxies
(Figure~\ref{fig:denbins}) the larger fractions for the highest density bin of field
galaxies (compared to cluster galaxies at similar densities) indicate the significance
of cluster environment related effects even in their outskirts 
\citep{bah13}. For bright field galaxies density is a key parameter for their 
evolution (although we cannot discard or distinguish internal feedback process). 
That is not true for faint field objects (Figure~\ref{fig:denbins2}), as low 
luminosity field galaxies are almost independent 
of environment. For cluster members the dependence on local density is very strong,
both for bright and faint galaxies. Considering galaxy stellar mass 
(Figures~\ref{fig:den_stmassbins} and~\ref{fig:rad_stmassbins}) we find the 
largest fractions of blue, star-forming and discs for lower mass galaxies.
Environment is not relevant for these objects until they are in regions with
local density typical of cluster outskirts. The environmental influence becomes
stronger for higher densities for these low mass galaxies. On the contrary, the 
most massive galaxies show small values of $F_B, F_L$ and $F_D$, as well as a 
smaller variation with density and cluster radius.

Comparing Figures~\ref{fig:denbins} and~\ref{fig:radbins} we see that bright 
cluster galaxies outside $R_{200}$ have local densities of about 2 galaxies Mpc$^{-2}$. 
In other words, outside the virial radius the local density drops very fast, 
which explains in part the strong transition we see once galaxies are within this
radius. Within the virial radius another transition is seen in the inner parts,
indicating that cluster related processes (tidal effects and ram pressure) further 
influence the galaxy populations. For faint galaxies, another characteristic region
is still seen at $> 1.8 \times R_{200}$. Using detailed simulations 
\citet{bah13} could investigate which processes are most common 
at these regions. They found that at $\sim 2-3 \times R_{200}$ {\em overshooting} and
pre-processing by groups explain the cold gas and star formation radial trends. But
to larger radius (reaching $5 \times R_{200}$) those two process can not fully explain
the hot gas radial trends, so that ram pressure is also relevant. Their results 
indicate that at large radius the ram pressure due to the ICM can strip the 
hot gas haloes both for low and high mass galaxies. The cold gas can be stripped 
only within $\sim R_{200}$.

\citet{kau04} show that even in low density regions the division of
galaxy populations in two broad families is valid. We corroborate these results
here, inspecting galaxies at different densities, but we also split them in 
cluster members and field galaxies. Even for the latter this division 
is still valid. When moving
to higher densities we can see the red, early-type sequence is well established with
a few transition galaxies still being noticed, such as bulges classified as star-forming
(or blue). That is also seen as function of cluster radius, with the red sequence
becoming narrower from the outer to inner regions of clusters. In agreement to
\citet{val11} we find that cluster mass is not a key parameter for
establishing the red sequence. But we find evidence that for high mass systems 
the red sequence show a reduced scatter compared 
to small groups. That result indicates that star formation is shut down and 
galaxies reach the red sequence already in the group scale, but the job is finished 
in higher mass systems, when the red sequence becomes tighter. Note again that 
\citet{val11} is restricted to the central region and, most 
important, to higher mass clusters.

Hence, our main results point to the variation of galaxy populations from very low
to the most dense regions of the Universe, with a stronger variation after galaxies
become part of groups/clusters (being found within $R_{200}$). The picture that 
comes out favors a pre-processing in the group scale, affecting first the
star-formation and in a second stage galaxy structure.
Our results indicate that local density 
is the main driver for galaxy evolution. The parent halo mass may be relevant only to the 
most central regions. However, that is a controversial issue in  the literature, 
possibly due to selection effects of the samples considered.
In agreement with \citet{bah13} and \citet{lu12}, we also find
an environmental dependence out to very large radius, corroborating the idea that
pre-processing by groups and {\em overshooting} act to transform galaxy properties.
The possible effect of ram pressure at large radius can only be tested using 
high resolution numerical simulations, as seen in the work 
of \citet{bah13}. Nonetheless, our main findings are in good 
agreement to what is found in semi-analytic models.

\section*{Acknowledgments}

ALBR thanks for the support of CNPq, grants 306870/2010-
0 and 478753/2010-1. PAAL thanks the support of CNPq, grant
304692/2011-5. We are also thankful to R. de Carvalho and F. La Barbera
for important suggestions regarding this manuscript.

This research has  made use of the SAO/NASA  Astrophysics Data System,
and the NASA/IPAC Extragalactic  Database (NED).  Funding for the SDSS
and  SDSS-II was  provided  by  the Alfred  P.  Sloan Foundation,  the
Participating  Institutions,  the  National  Science  Foundation,  the
U.S.  Department  of  Energy,   the  National  Aeronautics  and  Space
Administration, the  Japanese Monbukagakusho, the  Max Planck Society,
and  the Higher  Education  Funding  Council for  England.  A list  of
participating  institutions can  be obtained  from the  SDSS  Web Site
http://www.sdss.org/.



\label{lastpage}

\end{document}